\documentclass{article}

\usepackage{PRIMEarxiv}

\usepackage[utf8]{inputenc} % allow utf-8 input
\usepackage[T1]{fontenc}    % use 8-bit T1 fonts
\usepackage{hyperref}       % hyperlinks
\usepackage{url}            % simple URL typesetting
\usepackage{booktabs}       % professional-quality tables
\usepackage{amsfonts}       % blackboard math symbols
\usepackage{nicefrac}       % compact symbols for 1/2, etc.
\usepackage{microtype}      % microtypography
\usepackage{lipsum}
\usepackage{fancyhdr}       % header
\usepackage{graphicx}       % graphics
\graphicspath{{media/}}     % organize your images and other figures under media/ folder

\usepackage{multicol}

\usepackage{ifpdf}
\usepackage{cite}
\usepackage{amsmath}
\usepackage{algorithmic}
\usepackage{array}
\usepackage{url}
\usepackage{physics}
\usepackage{mathtools}
\usepackage{xcolor}
\usepackage{multirow}
\usepackage{textcomp}
\newcommand{\cent}{\textcent}
\usepackage[ruled, linesnumbered, vlined]{algorithm2e}
\allowdisplaybreaks

\title{Robo-Chargers: Optimal Operation and Planning of a Robotic Charging System to Alleviate Overstay
\thanks{Manuscript submitted to \textit{IEEE Transactions on Smart Grid} for review on December 6, 2022. 
Revised on March 15 and June 7, 2023. Accepted on June 11, 2023.
DOI: \href{https://doi.org/10.1109/tsg.2023.3286434}{10.1109/TSG.2023.3286434}}.
\thanks{\emph{Corresponding author}: Scott Moura.
This work was supported in part by TotalEnergies.}
}

\author{
  Yi Ju \\
  Center for the Built Environment \\
  University of California, Berkeley \\
  Berkeley, CA, USA\\
  \texttt{juy16thu@berkeley.edu} \\
   \And
  Teng Zeng \\
  Department of Civil and Environmental Engineering \\
  University of California, Berkeley \\
  Berkeley, CA, USA\\
  \texttt{tengzeng@berkeley.edu} \\
  \AND
  Zaid Allybokus \\
  TotalEnergies OneTech \\
  92078 Paris La Défense Cedex, France \\
  \texttt{zaid.allybokus@totalenergies.com} \\
  \And
  Scott Moura \\
  Department of Civil and Environmental Engineering \\
  University of California, Berkeley \\
  Berkeley, CA, USA\\
  \texttt{smoura@berkeley.edu} \\
}

\begin{document}
\maketitle

% keywords can be removed

% As a general rule, do not put math, special symbols or citations
% in the abstract or keywords.
\begin{abstract}
Charging infrastructure availability is a major concern for plug-in electric vehicle users. Nowadays, the limited public chargers are commonly occupied by vehicles which have already been fully charged. Such phenomenon, known as \emph{overstay}, hinders other vehicles' accessibility to charging resources. In this paper, we analyze a charging facility innovation to tackle the challenge of \emph{overstay}, leveraging the idea of \emph{Robo-chargers} - automated chargers that can rotate in a charging station and proactively plug or unplug plug-in electric vehicles. We formalize an operation model for stations incorporating Fixed-chargers and Robo-chargers. Optimal scheduling can be solved with the recognition of the combinatorial nature of vehicle-charger assignments, charging dynamics, and customer waiting behaviors. 
Then, with operation model nested, we develop a planning model to guide economical investment on both types of chargers so that the total cost of ownership is minimized. In the planning phase, it further considers charging demand variances and service capacity requirements. 
In this paper, we provide systematic techno-economical methods to evaluate if introducing Robo-chargers is beneficial given a specific application scenario. Comprehensive sensitivity analysis based on real-world data highlights the advantages of Robo-chargers, especially in a scenario where overstay is severe. Validations also suggest the tractability of operation model and robustness of planning results for real-time application under reasonable model mismatches, uncertainties and disturbances.
\end{abstract}

% Note that keywords are not normally used for peerreview papers.
% IEEE, IEEEtran, journal, \LaTeX, paper, template.

\section*{Nomenclature}
\newcommand{\ind}{\mathbf{1}}
\newcommand{\prob}{\mathbb{P}}
\newcommand{\R}{\mathbb{R}}
\newcommand{\Z}{\mathbb{Z}}
\newcommand{\N}{\mathbb{N}}
\newcommand{\B}{\{0,1\}}

\newcommand{\Tset}{\mathcal{T}}
\newcommand{\Iset}{\mathcal{I}}
\newcommand{\ta}{t^{\text{a}}_i}
\newcommand{\td}{t^{\text{d}}_i}
\newcommand{\einit}{E^{\text{init}}_i}
\newcommand{\etarg}{E^{\text{targ}}_i}
\newcommand{\edem}{E^{\text{dem}}_i}
\newcommand{\Xin}{\mathbb{I}_{i,t}}
\newcommand{\Xfix}{x^{\text{fix}}_i}
\newcommand{\Xrobo}{x^{\text{robo}}_i}
\newcommand{\Xleave}{x^{\text{leave}}_i}
\newcommand{\Xplug}{x^{\text{plug}}_{i,t}}
\newcommand{\vfix}[1]{v^{\text{fix}}_{#1}}
\newcommand{\vrobo}[1]{v^{\text{robo}}_{#1}}
\newcommand{\qfix}[1]{q^{\text{fix}}_{#1}}
\newcommand{\qrobo}[1]{q^{\text{robo}}_{#1}}
\newcommand{\Xqueue}{x^\text{queue}_{i,t}}

\newcommand{\Cost}[1]{C^{\text{#1}}}

\subsection*{Abbreviations}
% use full name at the first time it is explained, and abbr. later
% (not necessarily the first time it appears)
\textbf{PEV}: plug-in electric vehicle;
\textbf{FC}: Fixed-charger; 
\textbf{RC}: Robo-charger; 
\textbf{MCCS}: mixed-charger charging station; 
\textbf{FCS}: FC-only station; 
\textbf{SR}: satisfied rate;
\textbf{TOU}: time-of-use tariff;
\textbf{CAPEX}: capital expenditure;
\textbf{OPEX}: operation expenditure;
\textbf{TCO}: total cost of ownership;
\textbf{MPC}: model predictive control.

\subsection*{Indices / Sets}
\begin{itemize}
    \item $\Z$ for integers (non-negative $\Z_{\ge 0}$, positive $\N$). $\R$ for real numbers (non-negative $\R_{\ge 0}$, positive $\R_{> 0}$). $[a,b] \coloneqq \{x\in \Z \mid a \le x \le b\}$.
    \item $t, \Tset$: Index of time step within the optimization horizon and its set. $\Tset = [0,T]$ where $t=T$ is only defined for state variables.
    \item $i, \Iset$: Index of PEVs and its set. $\Iset = [1,I]$.
    \item $s, \mathcal{S}$: Index of typical profile scenarios and its set. $\mathcal{S} = [1,S]$. $\mathcal{I}_s = [1,I_s]$ is the set of PEV indices under scenario $s$ ($s$ is usually omitted when $S=1$).
\end{itemize}

\subsection*{Parameters}
\begin{itemize}
    \item $\Delta t$: Length of time steps.
    \item $M, N \in \Z_{\ge 0}$: Number of FC ports and RC ports.
    \item $\ta, \td$: Arrival and departure time of PEV $i$.
    \item $\Xin \in \B$: Whether PEV $i$ is (supposed to be) at the charging station at time $t$.
    \item $X^{\text{fix}}_{i,0}, X^{\text{robo}}_{i,0} \in \B$: Whether PEV $i$ has already been assigned to FC or RC before $t=0$.
    \item $\edem, \einit, \etarg$: The demand charge, initial charge and target charge of PEV $i$. $\edem = \etarg - \einit$.
    \item $\overline{P}_i$: Maximum charging power for PEV $i$.
    \item $P^\text{base}_t$: Base power of the charging station at time $t$.
    \item $\eta$: Charging efficiency.
    \item $\omega_i$: Waiting-tolerance factor of PEV $i$.
    \item $\rho$: Satisfied rate requirement.
    \item $\theta^{\text{sr}}$: Threshold for identifying "satisfied" sessions.
    \item $\gamma$: Charging fee for per unit of charge.
    \item $\beta_t$: TOU to import energy from the grid at time $t$.
    \item $\beta^{\text{fix}}, \beta^{\text{robo}}$: Capital cost of one FC or RC.
    \item $\beta^{\text{dc}}$: Demand charge per unit power.
    \item $\beta^{\text{switch}}$: Switch cost per charger plug-in and plug-out.
    \item $\beta^{\text{us}}_k, \theta^{\text{us}}_k$: Parameters defining a piece-wise linear penalty on short charge, where $\beta^{\text{us}}_k$ is unit penalty and $\theta^{\text{us}}_k$ is some threshold.
    \item $\pi_s$: Probability of sub-scenario $s$.
\end{itemize}

\subsection*{Decision variables}
\subsubsection*{Followings are main decision variables}
\begin{itemize}
    \item $m, n \in \Z_{\ge 0}$: Number of FC ports and RC ports to be optimized in the planning model.
    \item $\Xfix,\Xrobo,\Xleave \in \B$: Whether PEV $i$ is to be assigned to FC, RC, or leave directly.
    \item $\Xplug \in \B$: Whether PEV $i$ is plugged-in at time $t$.
    \item $p_{i,t}$: Charging power for PEV $i$ at time $t$.
    \item $e_{i,t}$: Charge of PEV $i$ at time $t$.
\end{itemize}
\subsubsection*{Followings are variables assisting the formalization}
\begin{itemize}
    \item $\widetilde{p}_{i,t}$: Curtailed charging power for PEV $i$ at time $t$.
    \item $\widetilde{e}_{i,t}$: Charge of PEV $i$ at time $t$ with curtailed power.
    \item $\qfix{i}, \qrobo{i} \in \Z_{\ge 0}$: Length of service queue of FCs and RCs at PEV $i$'s arrival.
    \item $\vfix{i}, \vrobo{i} \in \Z_{\ge 0}$: Number of service queue vacancies of FCs and RCs at PEV $i$'s arrival.
    \item $p^{\text{dc}}$: Maximum aggregate power in a billing cycle.
    \item $x^{\text{switch}}_{i,t} \in \B$: Whether PEV $i$'s plug-in status changes at $t$.
    \item $u^{\text{disapp}}_i$: Monetized penalty on PEV $i$'s disappointment.
    \item $r$: Satisfied rate.
\end{itemize}

\subsection*{Functions}
\begin{itemize}
    \item $\ind\{\cdot\}$: Indicator function. $\ind\{A\} = 1$ if $A$ is true else $0$.
    \item $[\cdot]^{+}$, $[\cdot]^{-}$: Positive and negative linear rectifier functions. $[x]^{+} \coloneqq \max\{x,0\}$, $[x]^{-} \coloneqq \min\{x,0\}$.
    \item $\lfloor \cdot \rfloor$: Floor function. $\lfloor x \rfloor \coloneqq \max \{z \in \Z \mid z \le x\} $
    \item $C(\cdot)$: Some cost function, e.g., $\Cost{TOU}$, $\Cost{disapp}$, $\Cost{dc}$, $\Cost{switch}$.
\end{itemize}

\section{Introduction}
\subsection{Background}

Plug-in electric vehicles (PEVs) become popularized over the past few years. Benefited from technology advancement and cost reductions \cite{gnann_what_2018}, the adoption rate is likely to continue increase, which contributes to reducing carbon emissions \cite{knobloch_net_2020}.
% The adoption rate growth is likely to continue under technology advancements and cost reductions \cite{gnann_what_2018}. Also, policymakers view the promotion as a strategic approach to reducing carbon emissions \cite{knobloch_net_2020}. 
Accessibility to charging facilities is among the top influential factors for PEV adoption \cite{gnann_what_2018}. 
Public charging infrastructure shows a positive causal effect on PEVs' market diffusion \cite{illmann_public_2020}, and also has potential to provide grid service \cite{powell_charging_2022}.
Public charging infrastructure helps resolve the ``range anxiety" of PEV owners, which shows a positive causal effect on PEVs' market diffusion \cite{illmann_public_2020}. Moreover, accessible public charging shows potential to provide grid service via frequency regulation and real-time ramping \cite{powell_charging_2022}. 
However, nowadays the number of public chargers is approximately one-tenth of on-road PEVs in the US \cite{muratori_rise_2021}. Worse still, many charging stations, especially those equipped with level-2 chargers, are commonly suffering from the overstay issue.

\emph{Overstay} is the phenomenon that a charger is occupied by a PEV after it has been fully charged \cite{zeng_inducing_2021}. Data analysis on a heavily-utilized charging station shows that, PEVs overstay for 1.5 hours on average in their charging sessions, approximately an extra of 75\% of the required charging time \cite{zeng_solving_2020}. Overstayed PEVs hinder others' accessibility to the relatively few charging resources, which is identified as a “bottleneck” of station service capacity \cite{lindgren_identifying_2015}.
A nationwide survey on Dutch PEV users shows a long-tailed distribution in session duration where 6\% of the most overstayed sessions occupy the charging facility for 30\% of the time \cite{wolbertus_fully_2018}.
The overstay issue also hampers stations ability to achieve higher revenues by serving more PEVs. 

In this paper, we propose an innovative solution to incorporate \emph{Robo-chargers} to alleviate overstay and enhance stations’ service capability.

\subsection{Related works}

In literature review, we concentrate on literature targeting the overstay issue. Existing approaches can be roughly categorized into three directions: (i) infrastructure upgrades; (ii) penalty or incentive design to regulate overstay behaviors; (iii) “interchange” within charging sessions.

\begin{table*}[!ht]
\centering
\caption{Related literature summary.}
\label{tab:my-table}
\resizebox{\textwidth}{!}{%
\begin{tabular}{m{0.2\textwidth} m{0.12\textwidth} m{0.25\textwidth} m{0.32\textwidth}}
\toprule
\textbf{Overstay solutions} & \textbf{Refs} & \textbf{Applications} & \textbf{Limitations} \\ \midrule
\textbf{Infrastructure upgrades} &
    \cite{lindgren_identifying_2015, graber_two-stage_2020, zhang_optimal_2017, dong_planning_2016, sarker_optimal_2015}
   &
  Multi-cable chargers, fast chargers, battery swapping, etc. &
  High capital cost; low utilization at non-peak hours; high demand charge \& transformer upgrades \\
\textbf{Pricing strategy} &
    \cite{biswas_managing_2016, zeng_inducing_2021, lu_deadline_2022, ghosh_menu-based_2018, bitar_deadline_2017, xu_planning_2018}
   &
  Penalize overstay; incentive flexible schedules, etc. &
  Lack empirical data; insensitivity in price; computational challenge \\
\textbf{Interchange management} &
    \cite{zeng_solving_2020, qiu_charging-as--service_2021, lai_-demand_2022, huang_design_2015, saboori_mobile_2022, zhang_mobile_2020}
   &
  Human valet; vehicle-to-vehicle (V2V); charging-as-service, etc &
  Station-level energy management is overlooked; lack holistic model for planning \\ \bottomrule
\end{tabular}%
}
\end{table*}

\begin{figure*}[h]
    \centering
    \includegraphics[width=18cm]{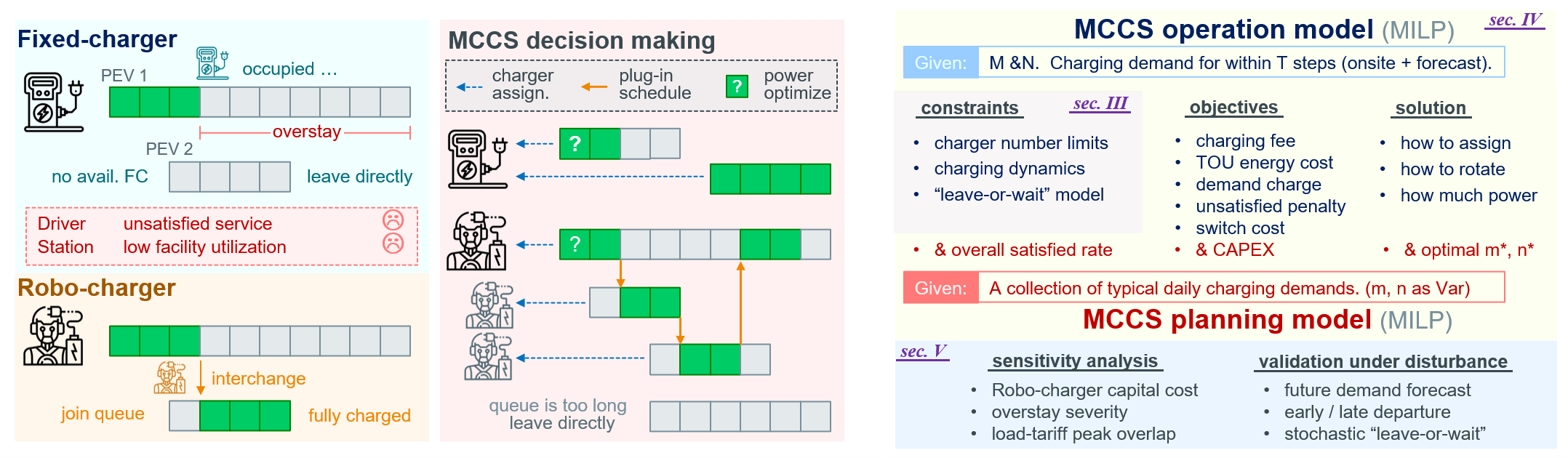}
    \caption{Overview of our work.
    \emph{left}: conceptual analysis of FCs \& RCs and MCCS decision model. \emph{right}: structure of this paper.
}
    \label{fig:concept}
\end{figure*}

Infrastructure upgrades, such as installing more chargers \cite{graber_two-stage_2020} (including multi-cable chargers \cite{lindgren_identifying_2015}, \cite{zhang_optimal_2017}), promoting fast chargers \cite{dong_planning_2016}, or shifting to battery swapping \cite{sarker_optimal_2015}, can improve service capacity of a charging station / network. However, such upgrades largely increase investment in the early stage. High aggregate power results in transformer upgrades and high demand charge, which are substantial costs overlooked in many studies \cite{he_fast-charging_2019, zhang_real-time_2018}.
Shifting charging mode to battery swapping would be a revolution across the entire PEV industry and there’s no evidence that it will become a dominant method.

Price menu design of charging fees is an important component of charging station management, and remains an active area of research. There are generally two ideas: One is to urge PEVs to leave as soon as possible by introducing an hourly overstay penalty \cite{biswas_managing_2016}. Alternatively, if overstayed PEVs accept flexible charging schedules, they can be managed as controllable loads, thus stations incentivize such choices\cite{zeng_inducing_2021, xu_planning_2018}. 
% Thus, some research works on the incentive design to encourage customers to accept a more flexible charging schedule \cite{zeng_inducing_2021}. Some recent works propose more detailed frameworks on designing deadline-differentiated pricing menus \cite{bitar_deadline_2017}, including considering heterogeneous drivers \cite{ghosh_menu-based_2018} and incomplete information \cite{lu_deadline_2022}. 
The specific price menu design can be highly complicated given the heterogeneity and stochasticity in behavioral patterns and dynamics in equilibrium \cite{bitar_deadline_2017, ghosh_menu-based_2018, lu_deadline_2022}. However, their real performance is largely remain untested since high-quality empirical data on customers’ behavioral model is very limited \cite{huang_are_2021}. Also, some research show that people are not quite sensitive to cost saving when comparing with increasing inconvenience and uncertainty.

“Interchange” basically means a charger may unplug a plugged PEV within its duration, and rotate to another PEV in need. “Interchange” is proposed in \cite{zeng_solving_2020} where the authors analyze the balance between initial investment of more chargers and operational cost of more interchanges (assuming done by human valets). Some related works apply this idea to district networks accompanied with mobile chargers (for instance, using vehicles to charge other vehicles, aka V2V \cite{qiu_charging-as--service_2021}) or employing human couriers \cite{lai_-demand_2022}. Such operations are actively discussed from the perspectives of optimal facility sizing \cite{huang_design_2015}, efficient routing algorithms \cite{qiu_charging-as--service_2021}, power grid benefits \cite{saboori_mobile_2022}, reservation coordinations \cite{zhang_mobile_2020}, market equilibrium analysis \cite{lai_-demand_2022}, en-route charging service \cite{qiu_charging-as--service_2021} and emergency management, etc.

Our proposed approach, so-called \emph{Robo-chargers}, is at the intersection of the three streams above. It is a charging facility innovation, yet also innovations to station operations. It enhances station's service capacity at peak hours, while also improves the overall charger utilization. The model incorporates drivers' queuing behavior based on quite simple and natural assumptions. Before diving into formalization details of Robo-chargers, some remarks on the aforementioned research from a methodological perspective are as follows:

\begin{itemize}
    \item
    We highlight the combinatorial optimization nature of optimal charging problem (with interchanges) given the limited number of chargers. However, few studies (e.g., \cite{liu_optimal_2021}) explicitly formalize it. Others assume unlimited charger accessibility, or simply relax the charging power limit of each charger into an aggregate version \cite{zeng_solving_2020}, \cite{woo_pareto_2021}.
    \item
    A stationary analysis for site planning, usually derived from time-invariant queuing theory, is commonly adopted \cite{lai_-demand_2022}, \cite{huang_design_2015}, \cite{zhang_mobile_2020}. However, a precise model of actual demand patterns (which is highly time-variant) does matter since station congestion has a bottleneck effect \cite{lindgren_identifying_2015}. Another limitation of such models is that charging power optimization for each session cannot be easily integrated. As a consequence a great opportunity for cost reduction (especially when the station is lightly loaded) is missed.
    \item
    In existing literature, customers’ waiting tolerance is modeled as a time threshold with the assumption that drivers have to wait in their cars \cite{qiu_charging-as--service_2021}, \cite{zhang_mobile_2020}, \cite{liu_optimal_2021}. With Robo-chargers, drivers can park their cars in the station and leave to do their own business. In such context, a new decision pattern based on expected charge at scheduled departures is formalized, and we analyze the chance of unsatisfied charging.
    % their decisions are based on whether their PEVs get fully charged at departure. Besides, there is no guarantee that all sessions can be exactly satisfied on time for a crowded station. We find discounts for non fully-charged batteries in a battery swapping station provides a referential formalization \cite{sarker_optimal_2015}.
    \item
    Interchange / mobile-charger’s application at the network level is seemingly more extensively discussed than at the single station level. However, to solve the overstay issue (the purpose of our paper), it is sufficient to manage them at a single station level. Reallocating chargers in the network is only needed when considerable spatially heterogeneous charging demand fluctuations exist.% (e.g., some stations are heavily loaded in the morning while some others have their peak hour in the afternoon).
\end{itemize}

Lastly, either planning or real-time scheduling requires a forecast module to predict the charging demands (as either discrete events or continuous arrival rates). Emerging machine learning algorithms contribute to the task \cite{al-ogaili_review_2019},\cite{ yi_electric_2021},\cite{xu_planning_2018} while it largely remains an open problem for further explorations. A comprehensive comparison of state-of-the-art practices is out of our scope. Instead, we demonstrate that with receding horizon control, a quite simple forecasting model can perform reasonably well.

\subsection{Contributions}

In this paper, we propose optimization models for station operation and facility planning with Robo-chargers. A summary of contributions is listed as follows:

\begin{enumerate}
    \item
    A conceptual model of \emph{Robo-chargers} is proposed. Robo-chargers can proactively rotate among PEVs for charging service, and help alleviate the overstay issue;
    \item
    An optimal operation model is formalized for rigorous management of charger assignment, plug-in schedules and power optimization. The model can be reformulated as mixed-integer linear programming (MILP);
    \item
    A planning model which optimizes the combination of Fixed-chargers and Robo-chargers is developed with the operation model nested. The model further incorporates realistic considerations such as weekly/seasonal charging demand variances and service capacity requirements;
    \item
    Sensitivity analysis suggests upgrading FC-only stations to MCCS is advantageous under a variety of scenarios. The advantage is also robust under potential uncertainties and disturbances with model predictive control implemented.
\end{enumerate}

We recognize that the full-stack engineering to make MCCS come true requires both operation research at the \emph{energy management} level, and detailed implementations on each \emph{motion-planning} tasks. We focus on the former level, and suggest that our model is not specific to any of various available modularized technical solutions for the latter level. Please refer to Appx.~\ref{sec:hardware} for more details on the hardware requirements.

The rest of the manuscript is organized as follows:
Conceptual analysis on Robo-chargers and mixed-charger charging stations is presented in  Sec.\,\ref{sec:concept}, which also serves as an outline for detailed mathematical formulations in Sec.\,\ref{sec:constr} (constraints) and Sec.\,\ref{sec:opt_formulation} (models). In Sec.\,\ref{sec:res}, we present numerical studies, including sensitivity analysis and uncertainty analysis. The manuscript is concluded in Sec.\,\ref{sec:conclusion}.

\section{Conceptual Analysis of Robo-chargers Planning Problem}
\label{sec:concept}
In this section, we formally define \emph{Robo-chargers} (RCs), in comparison to conventional \emph{Fixed-chargers} (FCs). Then, we introduce the highlights of our proposed operation and planning models for \emph{mixed-charger charging stations} (MCCS).

\subsection{Robo-chargers \& Fixed-chargers}
Today’s off-the-shelf FCs will be occupied by the parked PEVs throughout their plug-in duration, no matter if they have already been fully-charged or not. On the contrary, RCs are robot-based chargers that can automatically plug, unplug, and move to PEVs (illustrated in Fig.\,\ref{fig:concept}). 

An obvious operation improvement is that RCs can unplug PEVs once they have been fully-charged, and be available for others, thus the station no longer suffer from the opportunity loss caused by overstay. Another observation is that, waiting would be much more acceptable with RCs, since drivers can add their cars in the waiting queue of RCs by simply picking a spot to park them and informing the system of their departure times. Drivers can then leave to do their business. Meanwhile, RCs will strategically rotate between vehicles, performing interchanges based on optimized schedules.
% to provide charging service. They can even pause service to one vehicle and return later, depending on demand.

The observations above provide intuitions for the advantages of RCs. Meanwhile, several concerns arise. First, since RCs are more advanced in both hardware and software than FCs, their capital cost is higher. The trade-off between increased initial investment and improved operation revenues needs to be carefully balanced. Second, meeting customer expectations requires highly optimized dispatch operations, and there is usually no guarantee for waiting vehicles. If performing sub-optimally, some of the waiting cars may not get fully-charged, which would result in disappointment. 

\subsection{Optimization models for MCCS}

We construct a general model to include both FCs and RCs, known as the mixed-charger charging station (MCCS) model. A two-stage optimization model is developed: 
\begin{itemize}
    \item \emph{operation model}: Given the number of FCs and RCs, optimize the station's operation to minimize operation expenditure (OPEX).
    \item \emph{planning model}: With operation model nested, further determine the optimal number of FCs and RCs for minimization of total cost of ownership (TCO), which is the sum of OPEX and capital expenditure (CAPEX).
\end{itemize}
We are going to formally state the models in Sec.\,\ref{sec:constr} \& \ref{sec:opt_formulation}, while an outline here (also see Fig.~\ref{fig:concept}) may better navigate the readers to the core ideas.

The MCCS operation model includes three types of decision making: (\emph{i}) For each serviced PEV we either assign it to a specific FC, or add it into the service queue of RCs. (\emph{ii}) At each time step, RCs decide how to connect with PEVs in the service queue, if there are more PEVs than RCs; (\emph{iii}) Optimize charging power of each charger at each time step.

We enforce the constraint that only a limited number of chargers can be accessed simultaneously. We model drivers’ \emph{leave-or-wait} behavior and integrate it into the optimization model. We consider the charging dynamics and physical constraints, and consider time-of-use (TOU) tariff, demand charge, as well as a penalty for unsatisfied charging service.

When extending the operation model to a planning model, we treat the number of FCs and RCs as variables. Most constraints and objective terms are inherited, but some more are added: CAPEX is included in the objective; satisfied rate and daily load profile variances can be considered.

\section{Formalization of Operation Constraints}
\label{sec:constr}

This section details three groups of operation constraints.

\subsection{Characteristics of Fixed-chargers and Robo-chargers}

A potential charging session of PEV $i$ is characterized by a three-tuple $(\ta, \td, \edem)$ of its arrival time, departure time and energy demand. We denote $\Xin \coloneqq \ind\{\ta \le t < \td\}$ to indicate whether PEV $i$ is in the charging station at time slot $t$. 
Upon arrival, each vehicle will decide to whether stay and take the service at the charging station ($\Xleave=0$), or leave directly ($\Xleave=1$). For those choosing to stay, they are assigned to either FCs ($\Xfix=1$) or RCs ($\Xrobo=1$), so
\begin{equation}\label{eq:status_sum}
    \Xfix + \Xrobo + \Xleave = 1,\;\;\;\;\forall i
\end{equation}

When PEV $i$ is not supposed to be at charging station at time $t$, or it chooses to leave directly, it is certainly not being plugged-in ($\Xplug=0$). When PEV $i$ is at the station, and if it is assigned to an FC, it is certainly being plugged-in ($\Xplug=1$); while if it is assigned to RCs, then its plug-in status can be time-varying and is to be optimized. Above rules can be written compactly as
\begin{equation} \label{eq:plug}
    \Xin \Xfix \le \Xplug \le \Xin \left(1-\Xleave\right),
            \;\;\;\;\forall i, \forall t
\end{equation}

Given the constraint the there are only limited number of chargers, the number of simultaneously plugged-in PEVs is limited accordingly. Suppose the charging station has $M$ FCs and $N$ RCs, then at any time $t$, there can be at most $M$ PEVs plugged-in to FCs and $N$ plugged-in to RCs, which reads:
\begin{align}
    & \sum_{i=1}^I {\Xfix \Xplug} \le M,
            \;\;\;\;\forall t
            \label{eq:less_than_M}\\
    & \sum_{i=1}^I {\Xrobo \Xplug} \le N,
        \;\;\;\;\forall t
        \label{eq:leass_than_N}
\end{align}
For FCs, \eqref{eq:less_than_M} also indicates that at most $M$ vehicles can be assigned to FCs simultaneously. While more than $N$ vehicles can be assigned to RCs simultaneously, but at most $N$ of them can be plugged-in. This is the key difference between the two types of chargers.

Technically, the following constraints should be considered:
\eqref{eq:n=0} explicitly requires $\Xrobo=0$ for all $i$ if $N=0$.
% (While for FCs, \eqref{eq:less_than_M} \& \eqref{eq:plug} together have already implicitly force this constraint.)
\eqref{eq:fix_init} and \eqref{eq:robo_init} indicate that for those already in service at the beginning of optimization horizon, their present charger types ($X^{\text{fix}}_{i,0}, X^{\text{robo}}_{i,0}$) will be kept.%\zaid{are X in Eqn. (6) and (7) introduced??}
\begin{align}
    & \Xrobo \le N,
        \;\;\;\;\forall i
        \label{eq:n=0}\\
    & \Xfix \ge X^{\text{fix}}_{i,0},
        \;\;\;\;\forall i
        \label{eq:fix_init}\\
    & \Xrobo \ge X^{\text{robo}}_{i,0},
        \;\;\;\;\forall i
        \label{eq:robo_init}
\end{align}

\subsection{Charging schedules and PEV charge status}

For each PEV, the following constraints for charging power and status of energy (SoE) should be satisfied\footnote{Protocols for Energy Internet, e.g., ISO/IEC/IEEE 18881, enables secure information exchange between PEVs and the station \cite{wang_survey_2017}.}:
\begin{align}
    & 0 \le p_{i,t} \le \overline{P}_i \Xplug,
            \;\;\;\;\forall i, \forall t
            \label{eq_n:p}\\
    & e_{i,t} = e_{i,t-1} + \eta p_{i,t} \Delta t,
            \;\;\;\;\forall i, \forall t>0
            \label{eq_n:soc_update}\\
    & e_{i,0} = e_{i,\ta} = \einit,
            \;\;\;\;\forall i
            \label{eq_n:e_init}\\
    & e_{i,\td} = (\etarg-\einit)(1-\Xleave) + \einit,
        \;\;\;\;\forall i
        \label{eq_n:e_targ}
\end{align}
where $\edem = \etarg - \einit$.
\eqref{eq_n:e_targ} requires all PEVs get fully charged by their departures. However, in some circumstances, meeting such a target is either infeasible (e.g., too many waiting PEVs) or unprofitable (e.g., when the grid TOU is higher than charging revenue). We soften the constraint so that occasional violation is allowed, but the short in charge will be penalized in the objective. Technically, we introduce $\widetilde{p}_{i,t}$ to capture the curtailed power, which is eventually summed as the unsatisfied energy and be penalized. With $\widetilde{p}_{i,t}$ and the corresponding $\widetilde{e}_{i,t}$, constraints \eqref{eq_n:p} - \eqref{eq_n:e_targ} are modified as:
\begin{align}
    & 0 \le p_{i,t} \le \overline{P}_i \Xplug,
            \;\;\;\;\forall i, \forall t
            \label{eq:p}\\
   & e_{i,t} = e_{i,t-1} + \eta p_{i,t} \Delta t,
        \;\;\;\;\forall i, \forall t>0
        \label{eq:soc_update}\\
    & p_{i,t} \le p_{i,t} + \widetilde{p}_{i,t}
            \le \overline{P}_i \Xin,
        \;\;\;\;\forall i, \forall t
        \label{eq:p_tilde}\\
    & \widetilde{e}_{i,t} = \widetilde{e}_{i,t-1} + 
        \eta \left(p_{i,t}+\widetilde{p}_{i,t}\right) \Delta t,
        \;\;\;\;\forall i, \forall t>0
        \label{eq:soc_tilde_update}\\
    & e_{i,0} = \widetilde{e}_{i,0} = \einit,
        \;\;\;\;\forall i
        \label{eq:e_init}\\
    & \widetilde{e}_{i,\td} =
        (\etarg-\einit)(1-\Xleave) + \einit,
        \;\;\;\;\forall i
        \label{eq:e_targ}
\end{align}

\subsection{Behavioral model of leave-or-wait decisions}

PEV drivers decide whether to stay and wait for charging or leave immediately. However, their decisions depend on the charging station's operational situation upon arrival. Hence, although $\Xleave$ is a decision variable in the formulation, it (or more precisely, its probability distribution) can be determined given all the operations before $\ta$. We refer to the set of constraints determining $\Xleave$ as \emph{leave-or-wait} model. 

In general, drivers' decisions are based on their estimated chance that their PEVs would be fully charged by the declared departure times. However, an exact estimation would make the optimization problem intractable since station's operation model and drivers' decision model are deeply intertwined. Moreover, it does not make much sense to assume drivers would perform such complicated calculations in their mind before they make decisions. We adopt a simplified but more intuitive assumption that drivers' leave-or-wait decisions depend on the \emph{service queue length} at their arrivals. Here, service queue refers to all the PEVs in the station that would potentially compete for chargers. For FCs, it is simply all onsite PEVs assigned to FCs. For RCs, it refers to all PEVs in the station which are assigned to RCs \emph{and} have not been fully charged. 

Specifically, an \emph{$\omega$-tolerance} model is developed: Suppose there are $N$ RCs, the driver waits if there are available FCs, \emph{or} at most $\lfloor(1+\omega) N\rfloor-1$ PEVs are currently in the service queue of RCs. Otherwise it leaves. Let $\vfix{i}$ and $\vrobo{i}$ be the number of available vacancies (i.e., charging and waiting ports) for FCs and RCs at the arrival of PEV $i$, then above rule can be mathematically formalized as
\begin{equation} \label{eq:decide_leave}
    \Xleave = \ind \{\vfix{i} + \vrobo{i} \le 0\}
\end{equation}
Let $\qfix{i}$ and $\qrobo{i}$ be the queue lengths of FCs and RCs at the arrival of PEV $i$, then
\begin{align}
    & \vfix{i} = \left[M-\qfix{i}\right]^+,
    \;\;\;\;\forall i \label{eq:vfix}\\
    & \vrobo{i} = \left[\lfloor(1+\omega) N\rfloor -\qrobo{i}\right]^+,\;\;\;\;\forall i \label{eq:vrobo}
\end{align}
Suppose PEV indices are sorted by their arrival time, i.e., a smaller index indicates coming earlier thus also making the decision earlier (even though there are PEVs have the same arrival time). Then, $\qfix{i}$ and $\qrobo{i}$ follow the constraints
\begin{align}
    &\qfix{i} = \sum_{j=1}^{i-1} {x^{\text{fix}}_j \mathbb{I}_{j, \ta}},
        \;\;\;\;\forall i \label{eq:qfix}\\
    &\qrobo{i} = \sum_{j=1}^{i-1} {x^{\text{robo}}_j \mathbb{I}_{j, \ta} 
        \ind\{e_{j, \ta-1} < e^\text{targ}_j\}},
        \;\;\;\;\forall i \label{eq:qrobo}
\end{align}
Verbally, at the arrival of PEV $i$, FC queue includes all PEVs that arrive earlier, are assigned to FCs, \emph{and} still in the station by $\ta$. RCs queue includes those that arrive earlier, are assigned to RCs, still in the station, \emph{and} haven't been fully charged. It's easy to observe that the positive part function in \eqref{eq:vfix} can be omitted since $M-\qfix{i}$ is always nonnegative, but the positive part function in \eqref{eq:vrobo} is substantial.

A realistic extension of above model is to consider heterogeneous waiting tolerances, i.e., different drivers may have different thresholds to wait. Such extension can be made by simply replacing $\omega$ in \eqref{eq:vrobo} by $\omega_i$, where $\omega_i$ is the waiting tolerance of PEV $i$.\footnote{Abstract models known as bulk queue, reneging queue, retrial queue, etc., are investigated by queuing theory experts, and have implications for the real-world PEV charging systems. Though we adopt an atomic measure to model each individual session explicitly, and address relevant issues more or less, the non-atomic perspectives may also be inspiring for interested readers \cite{varshosaz_day-ahead_2019}.}

\section{Charging Station Optimization Formulation and Reformulation}
\label{sec:opt_formulation}
In this section, we further formulate the operation and planning models as optimization problems.

\subsection{Optimal control model for MCCS operation}

Given the number of FCs $M$ and RCs $N$ and the charging demand $\mathcal{D}_{\Tset} \coloneqq \{(\ta,\td,\edem)\}_{i\in\Iset}$, as well as other parameters such as grid TOU $\{\beta_t\}_{t\in\Tset}$, charging power limits  $\overline{P}$ and waiting-tolerance factor $\omega$, MCCS management system seeks the optimal sequence of operations on:
\begin{enumerate}
    \item whether to assign a PEV to an FC or service queue of RCs $\left(\vb{x^{\text{fix}}}=\{\Xfix\}_{\Iset},~\vb{x^{\text{robo}}}=\{\Xrobo\}_{\Iset}\right)$;
    \item for PEVs assigned to RCs, when should they be plugged-in to get charged $\left(\vb{x^{\text{plug}}} = \{\Xplug\}_{\Iset \times \Tset}\right)$;
    \item when PEVs are being charged, what the charging power should be $\left(\vb{p}=\{p_{i,t}\}_{\Iset \times \Tset}\right)$.
\end{enumerate}

The objective is to maximize the operating profit, i.e., minimize the operating expenditure (OPEX) considering revenues from charging fee $\Cost{fee}$, expenses for grid energy imports $\Cost{TOU}$, penalty on customers' disappointment $\Cost{disapp}$, demand charges $\Cost{dc}$ and also switching costs $\Cost{switch}$.

\begin{align}
    \min_{\vb{x^{\text{fix}}}, \vb{x^{\text{robo}}}, \vb{{x^{\text{plug}}}}, \vb{p}} &\text{OPEX}\\
    = & \underbrace{\sum_{i=1}^I \sum_{t=0}^{T-1} {(\beta_t - \gamma) p_{i,t} \Delta t}}_{\mathclap{\Cost{TOU} - \Cost{fee}}}
        + \underbrace{\beta^{\text{dc}} p^{\text{dc}}}_{\mathclap{\Cost{dc}}} \notag\\
        & + \underbrace{\sum_{i=1}^I \sum_{t=0}^{T-1} {\beta^{\text{switch}} x^{\text{switch}}_{i,t}}}_{\mathclap{\Cost{switch}}}
        + \underbrace{\sum_{i=1}^I {u^{\text{disapp}}_i}}_{\mathclap{\Cost{disapp}}}\notag
\end{align}
\begin{align}
    &\text{subject to:} \notag\\
    &~~~~\text{constraints \eqref{eq:status_sum} - \eqref{eq:robo_init}, \eqref{eq:p} - \eqref{eq:qrobo}, and}\notag\\
    & p^{\text{dc}} \ge P^{\text{base}}_t + \sum_{i=1}^I p_{i,t},
    \;\;\;\; \forall t \label{eq:p_dc}\\
    & x^{\text{switch}}_{i,t} =
    \big\lvert x^{\text{plug}}_{i,t+1} - x^{\text{plug}}_{i,t} \big\rvert,
    \;\;\;\; \forall i, \forall t \label{eq:switch}\\
    & u^{\text{disapp}}_i = \left(1-\Xleave\right) \sum_{k=1}^K {\beta^{\text{us}}_k 
    \left[\theta^{\text{us}}_k \edem - (e_{i,\td}-\einit)\right]^{+}},
    \forall i \label{eq:u_disapp}
\end{align}
Demand charges are based on the maximum charging power within a billing cycle.
Switching cost $\Cost{switch}$ are included to penalize frequent plug-in and plug-out, thus avoiding some meaningless charging behaviors.
Customers' disappointment $u^{\text{disapp}}_i$ is evaluated by a piece-wise linear function of the unsatisfied charge of PEV $i$, where $\theta^\text{us}_k$'s are some thresholds for unsatisfied charge. Severe short in required energy (failing to meet smaller threshold $\theta^\text{us}_k$) will be more heavily penalized (larger $\beta^\text{us}_k$) than slight mismatch. It captures the diminishing marginal utility in a simple way \cite{sarker_optimal_2015}. 

\subsubsection{Model predictive control (MPC)}
When applying the operation model in real-world practice, uncertainties and disturbances, such as future charging demands, early or late departures, and/or stochastic waiting tolerance, should be taken into account. Model predictive control (MPC) is applicable to resolve such challenges \cite{jiao_online_2022}. MPC re-optimizes at each time step and can adaptively improve scheduling quality as more information becomes available. Meanwhile, within each step of optimization, model can be appropriately simplified at the horizon ``tail'', which provides opportunity to accelerate the programming process. Detailed algorithm for MPC is explained in Appx.~\ref{sec:MPC}.

\subsection{Optimal planning model for MCCS}
A market decision, such as whether to incorporating RCs in charging stations, depends not only on its OPEX, but also its capital expenditure (CAPEX). Herein, with the operation model nested in, we develop a planning model where the numbers of FCs and RCs are optimized. In the planning model, FC and RC numbers are treated as decision variables, denoted as $m$ and $n$, and the total cost of ownership (TCO), i.e., the sum of CAPEX and OPEX, is to be minimized. All constraints in the operation model can be inherited by simply replacing the given constant $M$, $N$ with decision variables $m$, $n$. 

\begin{align}
    \min_{m, n, \vb{x^{\text{fix}}}, \vb{x^{\text{robo}}}, \vb{{x^{\text{plug}}}}, \vb{p}} 
    &\text{TCO} = \text{OPEX} 
    + \underbrace{\beta^{\text{fix}} \cdot m + \beta^{\text{robo}} \cdot n}_{\text{CAPEX}}
\end{align}
\begin{align}
    &\text{subject to:} \notag\\
    &~~~~\text{constraints \eqref{eq:status_sum} - \eqref{eq:robo_init}, \eqref{eq:p} - \eqref{eq:qrobo}, \eqref{eq:p_dc} - \eqref{eq:u_disapp}}\notag\\
    &~~~~\text{with $M, N$ replaced by $m,n$}\notag
\end{align}

Besides, some extra constraints on the overall service quality and demand patterns can be added into the planning model.

\subsubsection{Satisfied rate (SR) requirement}
By minimizing the TCO of the charging station, we primarily treat the planning problem as pure commercial affairs. In scenarios where providing better charging service is unprofitable (e.g., the marginal expenditure to satisfy all demands is too high), stations may strategically reduce the number of chargers. However, as infrastructure construction, the benefit of charging accessibility is shared across the community. In other words, the externality should be somehow internalized in order to make a wise decision in the public welfare sense. Since comprehensively discuss the externality of charging station infrastructure is beyond the scope of our paper, we simply consider the case where a certain satisfied rate (SR) is required.

We define SR $r$ as the proportion of satisfied customers to all customers (including leaving), and enforce the constraint that SR is above some given requirement $\rho$:
\begin{equation} \label{eq:SR}
    r = \frac{1}{I} \sum_{i=1}^I {\ind 
        \big\{e_{i,\td} - \einit \ge \theta^{\text{sr}} \edem\big\}}
    \ge \rho
\end{equation}
where $\theta^{\text{sr}}$, e.g., $0.9$,  is some threshold that a session can be regarded as ``satisfied'' although the charging demand may not be exactly met. Constraint \eqref{eq:SR} can be added to the planning model so that the solved optimal charger numbers also meet the given SR requirement.

\subsubsection{Multiple typical sub-scenarios considered in planning}
Daily PEV charging demands fluctuate considerably across the year, so it helps to consider multiple typical daily demand profiles in planning. Suppose there are $S$ sub-scenarios to consider, indexed with $s = 1, \dots, S$ with corresponding probability $\pi_s$, the overall OPEX is the weighted average of OPEX under each sub-scenario.
\begin{equation}
    \text{OPEX} = \sum_{s=1}^S {\pi_s\, \text{OPEX}_s (\mathcal{D}_\Tset^s)}
\end{equation}
For constraints \eqref{eq:p_dc}, \eqref{eq:SR}, $S$ sub-scenarios are weighted-summed to form the new constraint. For other constraints, each is rewritten as $S$ individual sub-constraints, i.e., they hold for every $s$. 
%aforementioned, it is easy to be transferred into a multi-scenario case. 
% Some detailed discussions are provided in Appendix.~\ref{subsec:multi-scenarios}.

\subsubsection{Robustness of planning results} \label{sec:robust}
In the planning model, we assume MCCS management system has complete information on charging demands, as extracted in representative profiles. While in real-world application, there exist potential model mismatches, uncertainties and disturbances. Possible uncertainties include (1) long-term uncertainty, such as the PEV market growth in coming years; (2) short-term uncertainty, such as charging demand fluctuations in coming hours. 
% Besides, in the long term, the representative charging profile selection may lead to a biased TCO estimation. In the short term, unpredictable randomness exists, for example, drivers leave-or-wait decision making is actually stochastic, and they may also depart earlier or later than they declare.
We validate that our model are practical and robust under above possible uncertainties in Sec.~\ref{sec:uncertain}. 

% To verify that our models are practical for real-world applications, we validate their robustness under potential model mismatches, uncertainties and disturbances. We implement a stochastic virtual environment for simulation and deploy an MPC controller to manage the charging station. The MPC controller solves the same optimization problem as the operation model does, while only executing optimized operations for one step and re-scheduling at each time step.

% Possible uncertainties include (1) long-term uncertainty, such as the PEV market growth in coming years; (2) short-term uncertainty, such as charging demand fluctuations in coming hours. Besides, in the long term, the representative charging profile selection may lead to a biased TCO estimation. In the short term, unpredictable randomness exists, for example, drivers leave-or-wait decision making is actually stochastic, and they may also depart earlier or later than they declare.

% Taking all these into consideration, it is likely that the estimated TCO provided by the planning model is over-optimistic that may not actually be achieved in real applications. Moreover, although aforementioned uncertainties and disturbances exist in every environment, control algorithms of higher complexity are more prone to suffering from them. In other words, TCO estimation on MCCS is likely to be more biased than that on FCS. We want to validate that, (1) MCCS can robustly outperform FCS. (2) The optimal charger combination solved by the planning model is reasonably acceptable.

\subsection{Reformulation and solving programs}
Both the operation model and planning model can be reformulated as a mixed-integer linear programming (MILP) problem with some general techniques.
%summarized in Appendix.~\ref{subsec:linearization}.
We implemented the optimization models via \texttt{Gurobi Optimizer} in \texttt{Python}. Our base planning case can be solved in 3 - 5 minutes with MIP gap set as $1\%$ on a personal computer. \footnote{\texttt{Gurobi 9.5.2} with academic license. PC with Intel i7-9750H CPU @ 2.60GHz, 12 logical processors can be used.} When implemented as MPC, steps on the horizon tail are reasonably simplified. The tailored operation model can be solved in 10 seconds on average per 15-minute rescheduling step.

\section{Numerical Studies}
\label{sec:res}
% \textcolor{blue}{Here's the results.}
\newcommand{\Nref}{\textcolor{red}{[\text{ref?}]}}
% \yi{Use figures with low resolution for faster rendering (titled with "LR")}.

\begin{figure*}[h]
    \centering
    \includegraphics[width=16cm]{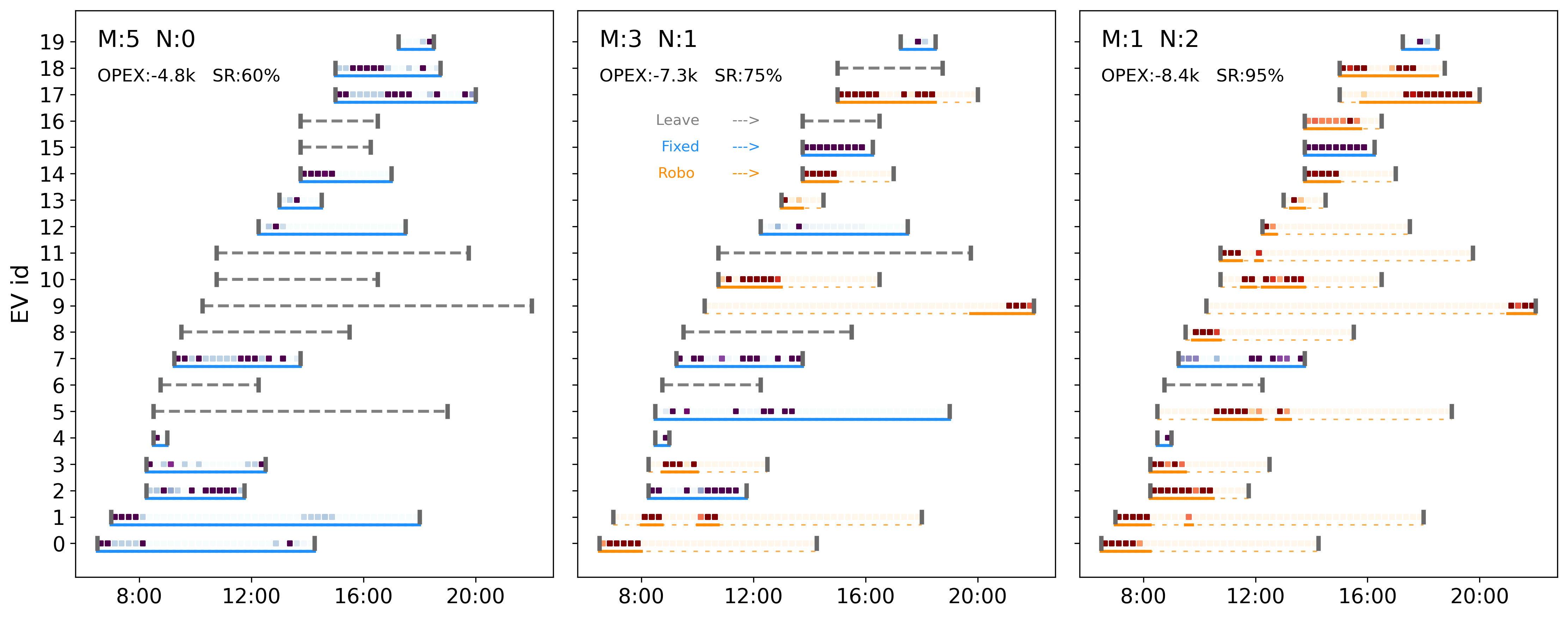}
    \caption[Caption for LOF]{Optimized charging station operations. 
    It is a conceptual illustration - a “toy” profile of 20 sessions is sampled as a case study here. 
    % \emph{Lower row}: 
    Each subplot represents an optimized operation schedule under corresponding charger combination settings, indicated by ``M'' and ``N'' at the left-top corner. Each horizontal bar represents one individual charging session. Gray dashed lines indicate leaving directly. Underline colors indicate charger types (blue-FC; orange-RC). Solid orange underline segments indicate being plugged-in by RCs. Shades of squares in the session bars represent charging power (deeper indicates greater power).
\textsuperscript{$\dagger$}}
  \footnotesize\textsuperscript{$\dagger$} A more detailed visual instruction on how to read this plot: \url{https://shorturl.at/pORV9}.
    \label{fig:operations}
\end{figure*}

In this section, we present simulations based on charging data from a real-world station. We demonstrate numerical results and visualizations of the operation and planning model. Further, we discuss the suitable scenarios for MCCS by a series of sensitivity analysis. Lastly, we validate the robustness of MCCS in real-world applications by uncertainty analysis.

\subsection{Data description and system configuration} \label{subsec:config}

\newcommand{\meanstd}[2]{{#1}_{\pm{#2}}}

PEV charging records from the Olser Parking Structure charging station on UCSD campus in 2019 \cite{silwal_open-source_2021} are used as the data source. The dataset includes detailed information of the start and end times and energy consumption for $12259$ charging events (daily $\meanstd{\text{mean}}{\text{std}}$: $\meanstd{33.6}{18.6}$).%\footnote{DC fast charging sessions are excluded.} 
Charging demands varied between weekdays ($\meanstd{42.9}{13.0}$) and weekends ($\meanstd{10.2}{4.4}$). We construct two typical demand profiles by randomly sampling $43$ and $10$ sessions from all the sessions on weekdays and weekends respectively. The sampled profiles are used as representative demands in planning.

The capital cost of FCs (including installment and maintenance fees) is estimated to be $\$\,5400$ each with a ten year lifespan via market survey \cite{nelder_reducing_2019}. We estimate the capital cost of a RC to be twice that of a FC in the base case, considering the complexity of its hardware, software, manufacture and maintenance. We also provide sensitivity analysis on RCs’ capital cost for reference. All chargers are level-2 chargers with maximum charging power of $6.6$ kW.\footnote{Our model itself is applicable for general chargers, AC or DC, level-1, 2, 3, etc., as long as its protocol allows controllable charging. However, in today's scenario, we consider it makes the most sense to update a level-2 station with RCs. Certainly, upper limits of charging power is an influential factor on both CAPEX and OPEX. Also, charging power upper limits, determined by both chargers (aka Electric Vehicle Supply Equipment (EVSE)) and PEV's onboard charger, can be heterogeneous.} PG\&E’s TOU plan for commercial charging stations is adopted, along with a $\$\,18\,/\,$kW demand charge fee per billing cycle (per month). Other parameters (for the base case) are summarized in Tab.~\ref{tab:params}.

% generate using: https://www.tablesgenerator.com/latex_tables
\begin{table}[h]
 \caption{Parameters used in base planning cases}
  \centering
\begin{tabular}{llll}
\toprule
\textbf{variable}   & \textbf{meaning}        & \textbf{value} & \textbf{unit}    \\ \midrule
$T$                 & optimization horizon        & $96$              &           \\
$\Delta t$          & step length                 & $0.25$            & hr        \\
$S$; $\pi$          & sub-scenarios num. \& prob. & $2$; $[5/7, 2/7]$ &           \\
$I$                 & session numbers             & $[43, 10]$        &           \\
$\omega$            & waiting tolerance factor    & $1$ \emph{or} $\infty$    &           \\
$\overline{P}$      & maximum charging power      & $6.6$             & kW        \\
$\gamma$            & charging fee                & $35$              & \cent/kWh \\
$\beta_t$           & TOU: super off-peak         & $11$              & \cent/kWh \\
                    & ~~~~off-peak TOU            & $13$              & \cent/kWh \\
                    & ~~~~peak TOU                & $34$              & \cent/kWh \\
$\beta^\text{dc}$   & demand charge (per month)   & $18$              & \$/kW     \\
$\beta^\text{us}$; $\theta^\text{us}$ & unsatified penalty params  & $[10, 20]$; $[1, 0.9]$ & \cent/kWh; \\
$\rho$; $\theta^{\text{sr}}$            & SR req. \& satified thres. & $0.9_{(\omega=\infty)}$; $0.9$           &            \\
$\beta^\text{fix}$  & capital cost of an FC       & $5400$            & \$        \\
$\beta^\text{robo}$ & capital cost of a RC        & $10800$           & \$ \\\bottomrule
\end{tabular}\label{tab:params}
\end{table}

We primarily consider two scenarios on the target customers: (1) customers have some alternative charging resources nearby so their waiting tolerance is low. Meanwhile, the charging station is purely profit-driven. In such a case, we assume $\omega=1$ for all PEV drivers and there is no SR constraint to enforce (the “$\omega=1$ case”). On the contrary, (2) customers will always stay to wait and accordingly the station should be planned to satisfy a given SR. In such a case, we assume $\omega=\infty$ and enforce a SR requirement at $90\%$ with $\theta^{\text{sr}}=0.9$ (the “$\omega=\infty$ case”).
It is worth mentioning that, though $\omega$ acts as an important empirical parameter to model customers’ behavior, its exact reference value can be very context-specific and is now unavailable. Our discussion on these two “representative” scenarios is a primitive attempt to investigate their influence categorically.

\subsection{Optimized charging operations with Robo-chargers}

As introduced above, for a given MCCS with determined number of FCs and RCs, the operation includes three types of decisions to maximize net profit, illustrated in Fig.\,\ref{fig:operations}: (1) For each arriving PEV (excluding those leaving directly), the station decides whether to assign it to an FC (blue underlines), or to add it into the service queue of RCs (orange underlines). (2) At each time slot, the station decides which PEVs in the service queue of RCs are plugged-in and receiving charge (solid orange underline segments). (3) At each time slot, the charging power of each charger is optimized (indicated by the shades of squared dots).

We can also clearly see the OPEX and SR differences across the three portfolios in Fig.\,\ref{fig:operations}, although they share the same CAPEX. This is further investigated in Sec.\,\ref{subsec:plan_res}.

% The three portfolios of chargers in Fig.\,\ref{fig:operations} share the same CAPEX. However, their OPEX and SR differ considerably. By replacing 2 FCs with 1 RC, both OPEX and SR improve even though the number of total chargers decreases. By transforming from “F5/R0”\footnote{Stands for 5 FCs and 0 RCs. Similar for others.} to “F1/R2”, the station revenue increases by \$ 3600 annually\footnote{Note that by definition, lower OPEX or TCO indicates better performance. Verbally, we use OPEX/TCO "improves", meaning it goes down. Graphically, in Fig.\,\ref{fig:sensitivity},\,\ref{fig:dgi},\,\ref{fig:mpc}, the lower the better. Values reported are always annualized.} and SR increases by 35\%.

\begin{figure}[h]
    \centering
    \includegraphics[width=8cm]{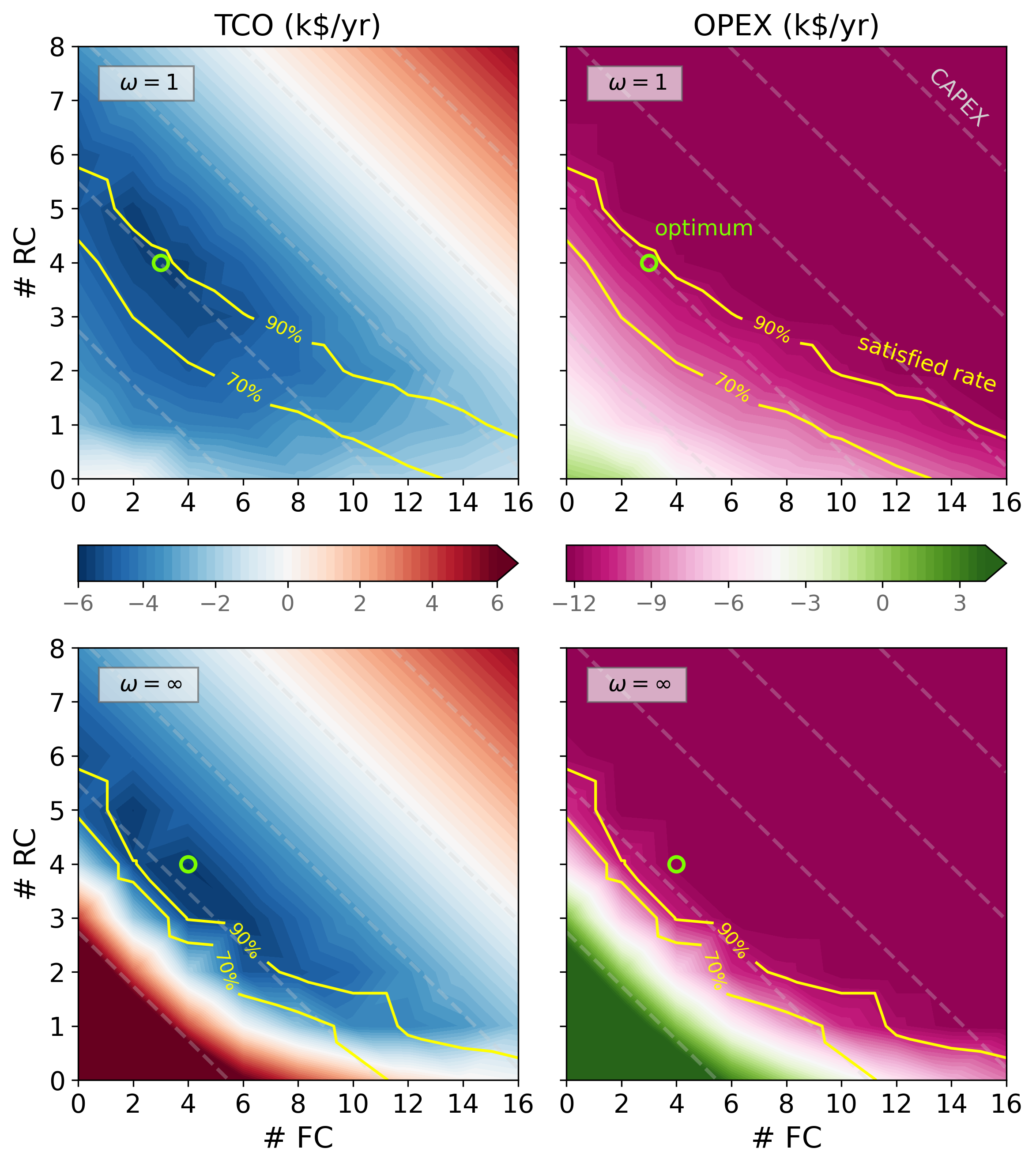}
    \caption{TCO \& OPEX of different charger combinations.
    Global optimum plans are marked in green circles. Heat maps of profits are generated based on grid search results of charger combinations. Dashed gray diagonal lines are CAPEX contours. Yellow lines are SR contours.
}
    \label{fig:planning}
\end{figure}

\subsection{Optimal investment plan for MCCS}\label{subsec:plan_res}

Considering the different characteristics of FCs and RCs, they have relative advantages under different circumstances. RCs are advantageous to stations that face severe overstay issues given their flexibility. Meanwhile, for sessions with little overstay, FCs are preferable for their lower capital costs. Given PEV charging demands as well as other required parameters, the planning model returns an optimal investment combination of FCs and RCs that minimizes the expected TCO. 

In our base case, the optimal plan is “F3\,/\,R4” for “$\omega=1$ case”, and “F4\,/\,R4”\footnote{Stands for 4 FCs and 4 RCs. Similar for others.} for “$\omega=\infty$ case”. The solved optima are marked in green circles in Fig.\,\ref{fig:planning}. Additionally, TCO and OPEX heat maps of varying charger combinations are included for comparison. The planning model can return a single optimal combination efficiently (solution time is in the same order of solving operation model once). The appended heat maps, exhaustively computed for each combination, also provide useful insights on, for example, where those sub-optimal solutions locate and how close their performances are.

Comparing “$\omega=1$ case” and “$\omega=\infty$ case”, more chargers are planned for the latter because customers always wait and $\text{SR}\ge90\%$ is enforced to satisfy. Meanwhile, more staying PEVs also provide opportunities to earn higher revenues. As a consequence, the $\omega=1$ case” yields an annualized net profit of $\$\,5233$ with (most economical) SR of $89.4 \%$, and the “$\omega=\infty$ case” yields $\$\,5611$ with SR $=100 \%$. 

Comparing optimal plan for MCCS and that for FCS,
%\footnote{Optimal plan for FC-only or RC-only stations can also be solved by our model, by simply adding a constraint $n=0$ (FC-only) or $m=0$ (RC-only).} 
in FCS, $8$ and $20$ FCs should be installed for “$\omega=1$ case” and “$\omega=\infty$ case” respectively. We find the main advantage of MCCS in ``$\omega=1$ case" is the OPEX improvement potential by serving more PEVs. While in “$\omega=\infty$ case”, MCCS is advantageous because much fewer chargers are required to meet a given SR, thus greatly reducing CAPEX.

\subsection{Sensitivity analysis}

The optimal investment plan is highly scenario-sensitive. We characterize the main influential factors as three variables: (1) Robo-charger capital cost ratio index (RCI); (2) Charging slackness index (CSI); and (3) Load-tariff peak overlap index (POI). We conduct a series of sensitivity analyses on them to provide references for practitioners. More importantly, it offers insights on the suitable conditions under which incorporating RCs can significantly improve the TCO of stations. 

\begin{figure}[h]
    \centering
    \includegraphics[width=8cm]{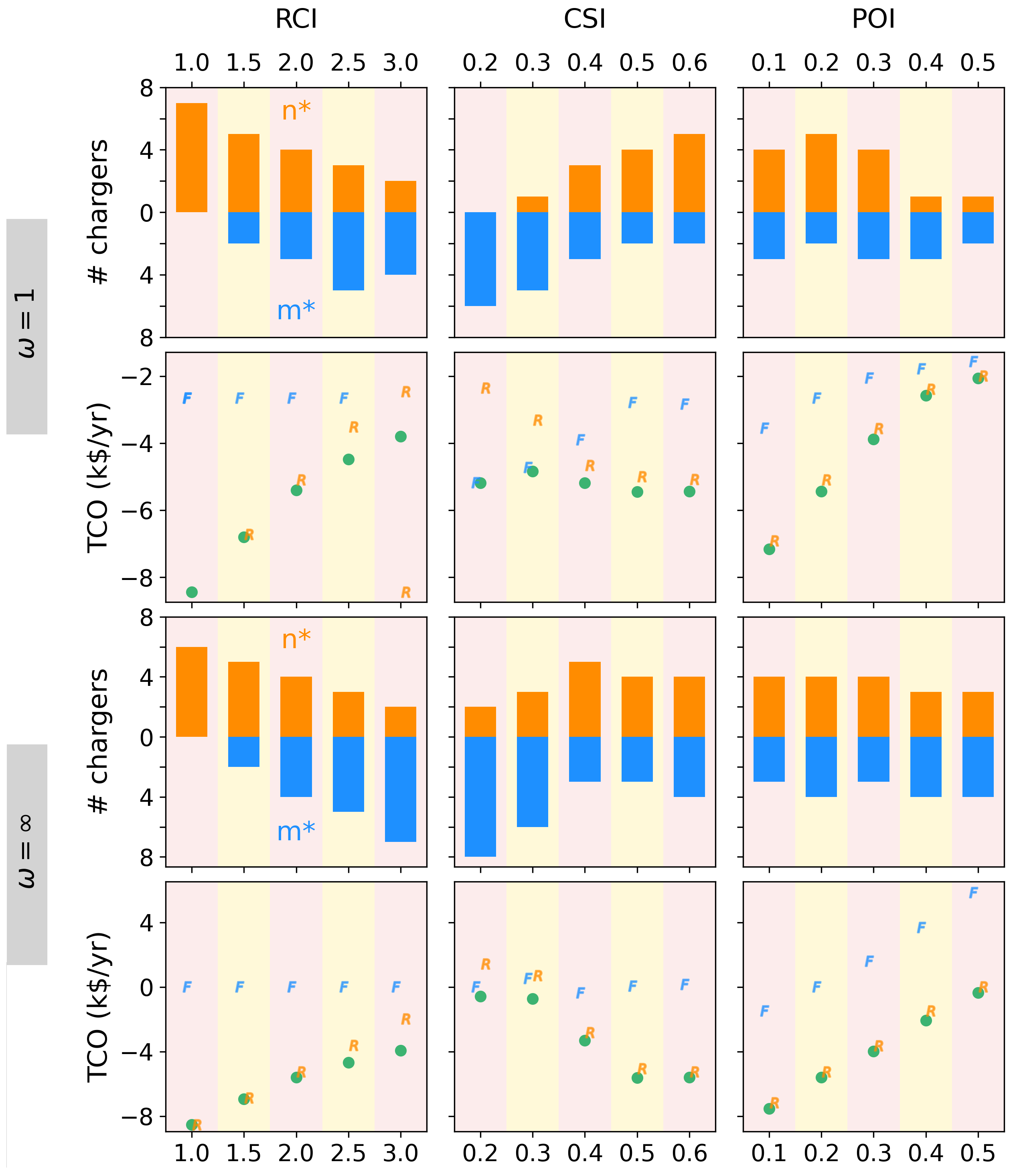}
    \caption{Sensitivity analysis. \emph{row 1\&3}: optimal planning results of charger numbers (blue: Fixed-chargers; orange: Robo-chargers). \emph{row 2\&4}: TCO corresponding to optimal plans. Green dots for global optimum. Blue markers for “only fixed-chargers” situation, while orange for “only Robo-chargers”.
}
    \label{fig:sensitivity}
\end{figure}

Figure.\,\ref{fig:sensitivity} visualizes the optimal planning results under different scenarios, including charger combinations and their corresponding TCO. TCO of optimally-planned FC-only stations and RC-only stations are also added for reference. For the base case: RCI $= 2.0$, CSI $=0.5$, POI $=0.2$.%\footnote{When varying CSI and POI, we generate some synthetic data based on the original sampled profiles. We manipulate each session consistently if possible, so that only the investigated variable is changed with other patterns kept. More details are introduced in corresponding paragraphs.} 

\subsubsection{Capital cost of Robo-chargers, RCI}
Since FCs are off-the-shelf products nowadays, their capital costs are relatively stable. However, estimation on RCs’ capital cost may have high variance at current stage due to the lack of in-depth design and manufacture details as well as survey on market willingness. We define \emph{Robo-charger capital cost index} (RCI) as the ratio of capital cost between one RC and one FC, i.e.,
\begin{equation}
    \text{RCI} = \beta^{\text{robo}}/\beta^{\text{fix}}
\end{equation}
with $\beta^{\text{fix}}=1.5$ \cent/day ($\$\,5400$ each for ten years) fixed.
The results are rather intuitive: since RCs can operate exactly the same as FCs (but not the reverse), when RCI $=1$, a RC-only plan will be the optimal. When RCI increases, the proportion of RCs in the optimal plan goes down.%\footnote{ The threshold beyond which the optimal plan will be FC-only is RCI $=6$ for “$\omega=1$ case” and RCI $=10$ for “$\omega=\infty$” case respectively (not shown in Fig.\,~\ref{fig:sensitivity}).}

\subsubsection{Charging slackness, CSI}
The severity of overstay is closely related to the need for RCs. We define \emph{charging slackness index} (CSI) to quantify it. Let \emph{slack} of a charging session $\widetilde{\tau}_i$ be the difference between PEV $i$'s duration at the station and the minimum required charging hours $\underline{\tau}_i$ (charging with power limits $\overline{P}_i$). Then CSI is defined as the average proportion of slack to its duration, i.e.
\begin{equation}
    \text{CSI} = \frac{1}{I}\sum_{i=1}^I{
    \frac{\widetilde{\tau}_i}{\widetilde{\tau}_i+\underline{\tau}_i}
    } 
    % = 
    % \frac{1}{I}\sum_{i=1}^I{
    % \frac{\left[\left(\td-\ta\right)\Delta t\right] - \left[\edem / (\eta \overline{P}_i)\right]}
    % {\left(\td-\ta\right)\Delta t}
    % }
\end{equation}
where $\underline{\tau}_i=\edem / (\eta \overline{P}_i)$ and $\widetilde{\tau}_i = \left(\td-\ta\right)\Delta t - \underline{\tau}_i$. 
Higher CSI indicates more severe overstay, where more OPEX improvement potential can be achieved by RCs. Another interpretation is, as CSI increases, the proportion of required charging time to the entire duration decreases. Thus, RCs can hold longer service queues without the concern of unsatisfied sessions.

\subsubsection{Load-tariff peak overlap, POI}
OPEX can be improved if most energy can be delivered during valley hours of the TOU plan, and vice versa. We define \emph{load-tariff peak overlap index} (POI) as the proportion of energy charged when TOU is at its peak values to all charged energy, assuming charging uniformly throughout duration, i.e.
\begin{equation}
    \text{POI} = \frac{1}{\sum_{i=1}^I{\edem}}
    \sum_{t=0}^{T-1}{\left[\ind\{\beta_t=\beta^{\text{peak}}\}
        \sum_{i=1}^I{\left(\Xin \frac{\edem}{\eta(\td-\ta)}\right)}\right]}
\end{equation}
As POI increases, more energy has to be charged at TOU peak. As a consequence, the average energy cost increases and makes some sessions less profitable. For “$\omega=1$ case”, the planning model strategically reduces charger numbers to save CAPEX, since OPEX increases because revenue does not offset costs as much.. For “$\omega=\infty$ case”, since a 90\% SR should be met anyway, charger numbers do not differ a lot, but TCO at high POI may even be positive, indicating the station will not be profitable by itself and subsidies are needed.

% this paragraph is removed (and replaced by discussion on long-term uncertainty in F.1)

% \paragraph{demand growth, DGI}
% Some regulations charge the stations for permission based on the number of chargers they install, regardless of fixed-chargers or the potential Robo-chargers. Under such circumstances, it is expected that the larger charging demands are, the higher proportion of Robo-chargers in all chargers should be. Demand growth index (DGI) is the ratio of daily session numbers to that of the base case (assuming they are sampled from the same distribution).

\subsection{Uncertainty Analysis}\label{sec:uncertain}

% \yi{I shrink this subsection to be of comparable length of other result parts - if you feel confused somewhere, please make a comment and you are welcome to refer to the complete version in this} \href{https://docs.google.com/document/d/14YKhMDntTNBauwGaA1GT97pNhjNJou5VrdGj2obx73M/edit?usp=sharing}{doc}

As motivated in Sec.~\ref{sec:robust}, since complete information and perfect execution is unreachable, it is likely that the estimated TCO provided by the planning model is over-optimistic and not achievable in real applications. We want to validate that, under potential long-tern and/or short-term uncertainties, at least: (1) MCCS can robustly outperform FCS. (2) The optimal charger combination solved by the planning model is acceptable.

% To verify that our models are practical for real-world applications, we validate their robustness under potential model mismatches, uncertainties and disturbances. We implement a stochastic virtual environment for simulation and deploy an MPC controller to manage the charging station. The MPC controller solves the same optimization problem as the operation model does, while only executing optimized operations for one step and re-scheduling at each time step.

% Possible uncertainties include (1) long-term uncertainty, such as the PEV market growth in coming years; (2) short-term uncertainty, such as charging demand fluctuations in coming hours. Besides, in the long term, the representative charging profile selection may lead to a biased TCO estimation. In the short term, unpredictable randomness exists, for example, drivers leave-or-wait decision making is actually stochastic, and they may also depart earlier or later than they declare.

% Taking all these into consideration, it is likely that the estimated TCO provided by the planning model is over-optimistic that may not actually be achieved in real applications. Moreover, although aforementioned uncertainties and disturbances exist in every environment, control algorithms of higher complexity are more prone to suffering from them. In other words, TCO estimation on MCCS is likely to be more biased than that on FCS. We want to validate that, (1) MCCS can robustly outperform FCS. (2) The optimal charger combination solved by the planning model is reasonably acceptable.

\begin{figure}[h]
    \centering
    \includegraphics[width=8cm]{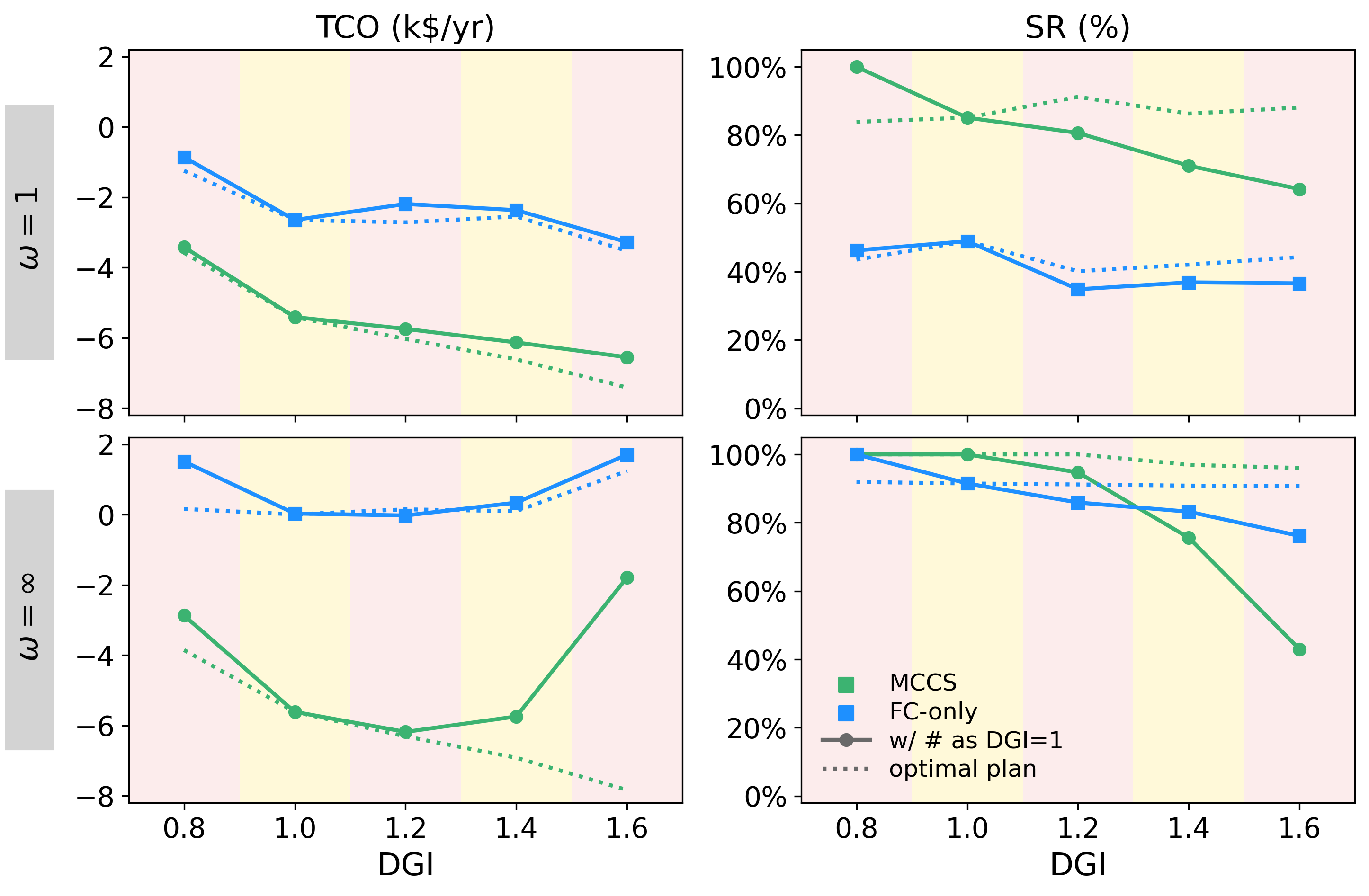}
    \caption{Sensitivity analysis under various DGI. \emph{left}: TCO. \emph{right}: SR. In each subplot, two plans - the optimal plan solved when DGI=1 (solid lines) for MCCS (green) or FCS (blue) - are tested under 5 DGI scenarios respectively. Dashed lines indicates TCO and SR if planned under the exact DGI.} % Note that a lower TCO is a better one. For w=inf case, all PEVs will wait for charging and their short in charge by departure will be penalized, even though there are no RCs.}
    \label{fig:dgi}
\end{figure}

\begin{figure}[!h]
    \centering
    \includegraphics[width=9cm]{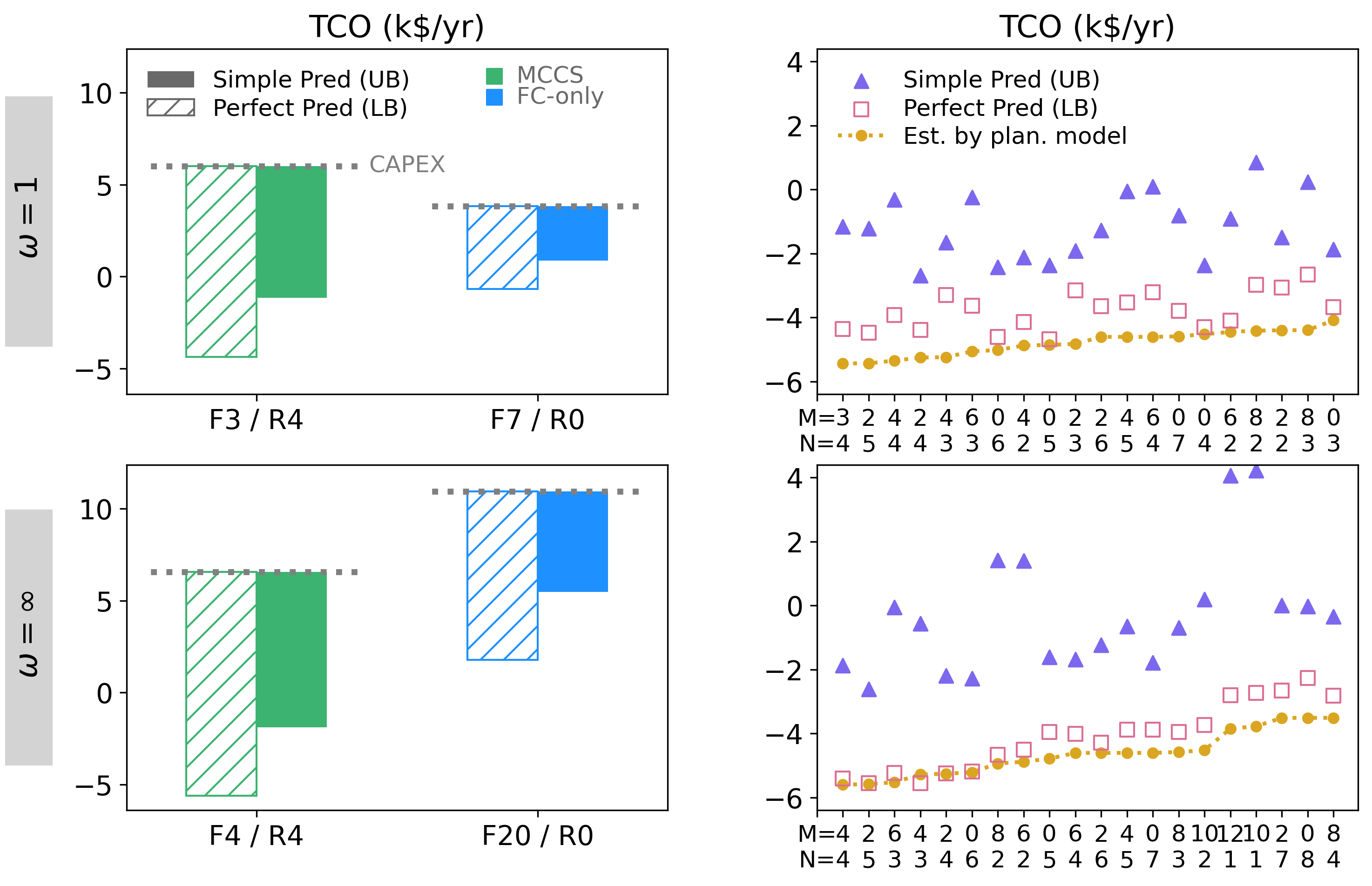}
    \caption{Simulated performances under short-term uncertainties and disturbances.
    \emph{left}: comparison between MCCS and FCS. Both are optimally invested as suggested by the planning model. Grey dashed lines (top of bars) mark CAPEX of the plans, and bottom of bars are TCO of the plans, so the lengths of bars indicate their OPEX. Fully-filled bars are under control with simple predictors (thus upper bound of TCO). Slash-hatched bars are under control with complete information (thus lower bound of TCO).
    \emph{right}: comparison among the top 20 combination candidates. Each dot represents TCO for one combination, ordered by their estimated TCO along $x$-axis. Gold circles are estimated TCO in the planning phase. Purple triangles and pink squares are simulated performances with simple predictors and complete information respectively.
    }
    \label{fig:mpc}
\end{figure}

\subsubsection{Long-term uncertainty}
For a specific charging station, it is difficult to foresee the localized future charging demand. We define \emph{demand growth index} (DGI) as the proportion of charging demands in a future scenario to that of today. We experiment with different DGI when the fixed “optimal” combination is solved at DGI$=1$. We compare both TCO and SR changes between MCCS and FCS, shown in Fig.\,\ref{fig:dgi}.

In all cases, even with up to 60\% of DGI underestimation, MCCS still outperforms FCS in terms of TCO. With increasing DGI, on the one hand, charger utilization may increase in times when demand was previously relatively low, thus earning more revenue. On the other hand, during peak hours, more PEVs either have to leave directly (if $\omega=1$) or do not meet the charge target by their departure (if $\omega=\infty$), thus SR decreases and the station may also get penalized. Consequently, in the ``$\omega=1$ case", growing DGI is generally beneficial to charging stations, even if they are undersized. While the ``$\omega=\infty$ case" is more sensitive to DGI. Lastly, considering that RCs are also more flexible to scale up and down, MCCS's advantages over FCS can be even larger (approaching the dashed lines).

% In the w=1 case, when DGI goes higher, both types of stations may achieve more profits from charging services. It is because chargers may not be fully utilized in some time slots (e.g., early morning or midnight) due to the daily variances in charging demands. When there are more demands, some of them fit in these unoccupied chargers. Meanwhile, SR decreases as DGI increases, because during peak hours more PEVs have to leave directly due to the long service queues.
% In the w=inf case, where all PEVs will park in the charging station no matter how long the service queue is, higher DGI may result in profit loss. It is because all waiting PEVs are expected to be fully-charged, otherwise the station will be penalized for customers’ dissatisfactions. However, since charger numbers are planned for DGI=1, they are not able to satisfy all sessions (otherwise such a plan is not economical when DGI=1). SR and DGI of MCCS drop more significantly than FCS, since chargers in MCCS are more efficiently used (especially, less likely to be unoccupied since there are waiting PEVs), and thus less room for extra sessions. It might be better to plan some redundant RCs when future demand is likely to increase considerably and the station has to meet some certain SR.

% However, in all cases, even with up to 60\% of DGI underestimation, MCCS still outperform FCS in terms of TCO. Moreover, RCs are also more flexible to scale up and down, so the advantages over FCS can be even larger.

\subsubsection{Short-term uncertainty}
We run simulations of station management with a MPC controller for one week (672 steps).\footnote{Details on how uncertainties are simulated can be found in Appx.~\ref{sec:stochasticity}.} To deal with unknown future demands, we set up a simple load forecast model that estimates and generates coming PEVs based on the hour in a day and whether the day is weekend or not. We also add random perturbations on waiting factors and departure times, which are assumed unpredictable for our simple predictor. The performance under such a predictor can be interpreted as the \emph{upper bound} %\zaid{is this THE upper bound or some pessimistic evaluation only? In this case please change the term}
(i.e., worst possible) of TCO in real applications, since a station can always develop such a predictor as long as it keeps the historical data, and there is plenty of room to improve. We set up another idealized controller which has complete information of ongoing and coming sessions for decision making. Its performance should be considered as the \emph{lower bound}  (i.e., best possible) of TCO.

In the left part of Fig.\,\ref{fig:mpc}, we compare TCO of MCCS and FCS. Suppose they are both controlled with simple predictors, or both with complete information, TCO of MCCS is always lower than FCS. Moreover, MCCS with simple predictors actually outperforms FCS with complete information, which suggests that the benefits of upgrading FCS to MCCS are guaranteed even under imperfect conditions.

In the right part of Fig.\,\ref{fig:mpc}, we further compare actual performances among top 20 charger combinations with the best estimated TCO. Ordered by their estimated TCO, their performances (the purple triangles) can be quite disordered, and some candidates outperform the chosen plan, i.e., the plan with lowest estimated TCO. It seems that plans with more chargers (high CAPEX) are more likely to show large TCO deviations, which might be because more coordinations are required for those plans, thus suffering more from incomplete information and randomness. However, all these ``elite" candidates are \emph{MCCS plans}, and difference in their estimated TCO is relatively small. So compared with FCS we have today, MCCS is hopefully to be a more profitable solution, and the MCCS plan suggested by our planning model would give a reasonable choice for station planning.

\subsection{Limitations}

Although the benefits of upgrading FCS to MCCS are supported by uncertainty analysis, we recognize that the results are not yet perfect. To our understanding, these deviations in estimated and actual performances are quite general challenges, but they have not drawn enough attention from the community. Some of our general thoughts are: 
On the one hand, improving the quality of forecasts can assist the system for better decision making. % Several works: \yi{xxx}. 
On the other hand, since perfect forecasts and zero disturbances are unreachable, integrating these considerations into the planning phase can be of substantial help. It leads to the active research area of robust optimization, but also more challenging in both formulation and computation. %We suggest interested readers to refer to works: \yi{xxx}. 
Lastly, as a human-in-the-loop societal system, there are also great needs to better understand customers’ behavioral patterns and design better market mechanisms accordingly. %Paper \yi{xxx} provide insights on this topic. 

\section{Conclusion}
\label{sec:conclusion}
We concentrate this paper on the conceptualization, formulation, simulation and result interpretation of the key characteristics of RCs and MCCS. We propose optimal operation and planning models for station management with RCs, which is suggested to be advantageous to today's FC-only stations. Interested practitioners can adopt our model for investment suggestions and implement the robust and efficient MPC algorithm for real-time operations. Moreover, our operation research on MCCS provides insights on how to better invest and utilize public charging infrastructures to attain a win-win outcome for PEV drivers, charging stations and power grid, which is a promising path towards a sustainable future.\footnote{Meanwhile, we also want to remind interested investors of several challenges from the hardware / motion-planning side. First, the compatibility of RCs and different makes of PEVs needs to be carefully considered. Second, the security of MCCS should be particularly emphasized, which may result in an increase in maintenance fees. Lastly, we assume the capital cost of RCs as some constant in our analysis, but as a brand new product, the high R\&D expense is non-neglectable, and thus the price strongly depends on how many RCs are sold.}

\appendix
\section*{Appendix}
\renewcommand{\thesubsection}{\Alph{subsection}}

\subsection{Hardware requirements}\label{sec:hardware}
\begin{figure}[!h]
    \centering
    \includegraphics[width=9cm]{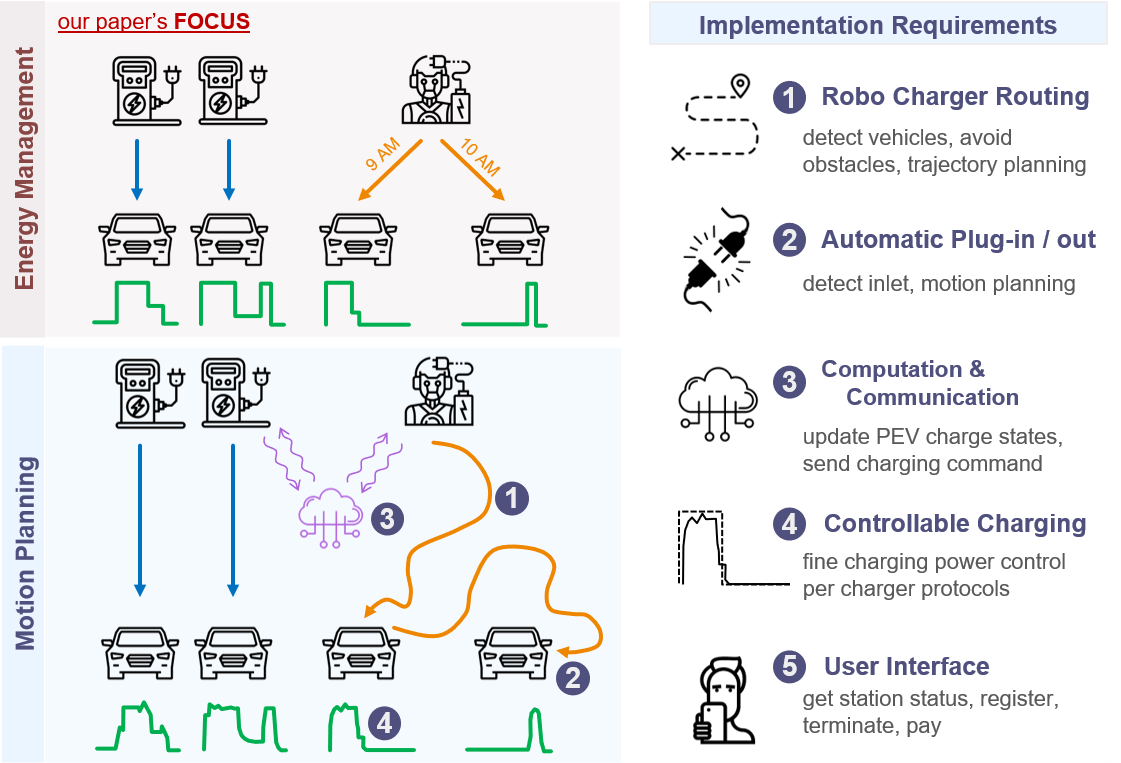}
    \caption{Hardware requirement to implement smart charging in MCCS.
    \emph{left}: a conceptual diagram highlights the differences between energy management and motion planning perspectives.
    \emph{right}: a summary of hardware (and also some software) requirements to implement our proposed system.
    }
    \label{fig:hardware}
\end{figure}

We provide a summary of the hardware (and also some software) requirements to implement our proposed system in the real-world in Fig.~\ref{fig:hardware}. It includes sensors and algorithms that enable Robo-chargers to detect vehicles, route to a target PEV, plug into / out from the PEV's charging port, communicate with a central server to update the PEV states, and receive charging schedules. Other infrastructure support may include a user interface (a machine in the station or a mobile phone application) that customers interact with to see charger occupancy, register their sessions, and make payments, etc.

The optimal operational model is not specific to any particular hardware or software. In fact, there are multiple hardware and software alternatives available in literature, for instance, vehicle recognition \cite{sochor_boxcars_2016}, routing \cite{shen_autonomous_2020}, automatic plug-in \cite{miseikis_3d_2017}, communication \cite{al-anbagi_wave_2016}, charging control \cite{mishra_sigma-modified_2022}, etc. Note the operational model focuses on the energy management level. The lower-level motion planning tasks are not a focus on this manuscript. 

\subsection{Model predictive control (MPC)}\label{sec:MPC}

\newcommand{\SetAttr}[2]{#1[#2]}

\newcommand{\taN}{t^{\text{a}}}
\newcommand{\tdN}{t^{\text{d}}}
\newcommand{\esttdN}{\hat{t}^{\text{d}}}
\newcommand{\edemN}{E^{\text{dem}}}
\newcommand{\einitN}{E^{\text{init}}}
\newcommand{\etargN}{E^{\text{targ}}}
\newcommand{\tildeeinit}{\widetilde{E}^{\text{init}}_{i}}
\newcommand{\tildeetarg}{\widetilde{E}^{\text{targ}}_{i}}
\newcommand{\ecurrN}{e^{\text{curr}}}
\newcommand{\pmaxN}{\overline{P}}
\newcommand{\XCN}{x^{\text{c}}}

\newcommand{\Dem}{\mathcal{D}_{\Tset}}
\newcommand{\Onsite}{\mathcal{L}}
\newcommand{\Vt}{\mathcal{V}_t}
\newcommand{\Vthat}{\widehat{\mathcal{V}}_t}
\newcommand{\Vttilde}{\widetilde{\mathcal{V}}_t}
\newcommand{\ecurr}{e^{\text{curr}}_i}
\newcommand{\esttd}{\hat{t}^{\text{d}}_i}
\newcommand{\XC}{x^{\text{c}}_i}
\newcommand{\Ot}{\mathcal{O}_t}
\newcommand{\Othat}{\widehat{\mathcal{O}}_t}
\newcommand{\Ottilde}{\widetilde{\mathcal{O}}_t}

Algorithm~\ref{alg:cap} provides a sketch of our MPC algorithm in the MCCS. Before optimization the scheme updates the system state from PEVs onsite, updates the future load forecasts, and considers uncertainties and disturbances revealed in the previous steps. Technically, the station maintains a data table tracking the status of all PEVs in the station, denoted as $\Onsite$. We use the notation $\SetAttr{\mathcal{A}}{k_i}$ to refer to values of field $k$ on index $i$ in data table $\mathcal{A}$.

\small
\begin{algorithm}
\caption{MPC algorithm for MCCS operation}\label{alg:cap}
% \SetKwInOut{Input}{Input}
% \SetKwInOut{Output}{Output}
\SetKwFunction{Pred}{GetPred}
\SetKwFunction{Opt}{SolveOp}
\SetKw{Init}{Initialize}
\SetKw{Exe}{Execute}

\KwIn{$\Dem$, $\Theta$, \Pred}
\KwOut{$\{\Ot\}_{t\in \Tset}$}

\Init $\Onsite$: an empty date table\;
\For{$t$ in $\Tset$}{
    \For{$i$ in $\Onsite$}{
        \lIf{$\SetAttr{\Dem}{\td}=t$}
            {remove $i$ from $\Onsite$}
    }
    \For{$i$ in $\Dem$}{
        \lIf{$\SetAttr{\Dem}{\ta}=t$ \& it stays}
            {add $\Dem[i]$ into $\Onsite$}
    }
    \Init $\Vt$: an empty data table\;
    \For{$i$ in $\Onsite$}{
        $\SetAttr{\Vt}{\ta} \gets 0$;~~~~
        $\SetAttr{\Vt}{\td} \gets 
            \max\{1, \Onsite[\esttd]-t\}$\;
        $\SetAttr{\Vt}{\einit} \gets \SetAttr{\Onsite}{\ecurr}$;~~~~
        $\SetAttr{\Vt}{\etarg} \gets 
            \SetAttr{\Onsite}{\etarg}$\;
        $\SetAttr{\Vt}{\omega_i} \gets \infty$;~
        $\SetAttr{\Vt}{\overline{P}_i} \gets \SetAttr{\Onsite}{\overline{P}_i}$;~
        $\SetAttr{\Vt}{X^{\text{c}}_0} \gets \SetAttr{\Onsite}{\XC}$
    }
    $\Vthat \gets$ \Pred{$t; \mathcal{D}_{\mathcal{T}^\prime}$}\;
    $\Vttilde \gets \text{merge}~\Vt, \Vthat$\;
    $\Ottilde = (\Ot, \Othat) \gets$ \Opt{$\Vttilde; \Theta$}\;
    \For{$i$ in $\Ot$}{ 
        \Exe $\SetAttr{\Ot}{p_{i,0}}$\;
        $\SetAttr{\Onsite}{\ecurr} \gets \SetAttr{\Ot}{e_{i,1}}$;
        % }
    }
}
\end{algorithm}

\emph{Nomenclature}
\begin{itemize}
    \item $\Dem$: Charging demands (PEV information) over the operation/simulation horizon $\Tset$ - indexed on PEVs' index $i$, with keys $[\taN, \esttdN, \tdN, \edemN, \omega, \pmaxN]$.
    \item $\Onsite$: A log keeping onsite PEV information - indexed on PEVs' index $i$, with keys $[\taN, \esttdN, \einitN, \etargN, \pmaxN, \XCN, \ecurrN]$, where:
    \begin{itemize}
        \item $\ecurr$: Current charge of PEV $i$.
        \item $\XC$: Charger type (FC or RC) PEV $i$ is assigned to (\texttt{NA} for new arrival PEVs).
    \end{itemize}
    \item $\Theta$: All the related parameters required in the operation model.
    \item \FuncSty{SolveOp}: Optimization solver for the operation model.
    \item \FuncSty{GetPred}: Predictor generating future charging instances.
    \item $\Vt$: Collection of onsite PEV information at time $t$ for \emph{operation model} \FuncSty{SolveOp}, including all parameters related to PEV. Similarly:
    \begin{itemize}
        \item $\Vthat$: information of predicted sessions
        \item $\Vttilde$:  concatenate $\Vt$ and $\Vthat$
    \end{itemize}
    \item $\Ottilde$: Optimized operations solved by \FuncSty{SolveOp}. Decompose into $\Ot$ and $\Othat$, where $\Ot$ are operations on onsite PEVs.
    % \item $\SetAttr{\mathcal{A}}{k_i}$: value of field $k$ on index $i$ in data table $\mathcal{A}$.
\end{itemize}

\emph{Remarks}
\begin{itemize}
    \item \emph{row 6} (\emph{``it stays''}):
    When a PEV arrives, a separate simulator will simulate if it will stay or leave directly. Details is described in Appx.~\ref{sec:stochasticity}.
    \item \emph{row 10}:
    We allow ``planned short in charge" ($\widetilde{p}_{i,t}$), however, $\ecurr$ does not track $\widetilde{p}_{i,t}$. To ensure the operation problem is always feasible, we adjust $\Vt[\etarg]$ to be $\min{\{\Onsite[\etarg], \Onsite[\ecurr] + (\Onsite[\esttd]-t) \overline{P}_i \eta \Delta t}\}$.
    \item \emph{row 11}:
    Penalty term on unsatisfied charge should always based on the original demand. To fix this, we include another two terms to the optimization problem: $\Vt[\tildeeinit] = \Dem[\einit], \Vt[\tildeetarg] = \Dem[\etarg]$ (not necessarily the same as $\Vt[\einit], \Vt[\etarg]$).
    
    \item For \emph{demand charge}, we include a parameter $\overline{p}^{\text{dc}}$ to track the maximum aggregate power observed in current billing cycle, and add $p^\text{dc} \ge \overline{p}^{\text{dc}}$ in constraint \eqref{eq:p_dc}.
    \item Strategies to \emph{accelerate the optimization}: (1) No need to consider PEVs which come later than all onsite PEVs have departed. (2) Introduce "varying intervals" to reduce decision variables. Instead of $15$-min intervals for $96$ steps, we consider $15$-min intervals for the first $8$ steps, $1$-hr intervals for the next $4$ steps, and $2$-hr intervals for the last $9$ steps ($21$ steps in total.)
\end{itemize}

\normalsize

\subsection{Stochasticity simulation}\label{sec:stochasticity}

We integrate three sources of uncertainties in the simulator to validate the robustness of our model in Sec.~\ref{sec:uncertain}.

\paragraph{Future charging demand} The optimizer does not know exact information on future sessions, which is used for solving the operation problem. We test the model with a naive forecast model, which has high forecast error. Namely, we extract two typical profiles from all historical sessions (randomly sampled from the number of daily average sessions), one for weekdays and one for weekends. Next, at every time step, the forecaster simply uses these averaged profiles for the future demand.

\paragraph{Heterogeneous and stochastic behavioral model} In the ``$\omega=1$ case'', the actual waiting tolerance factor $\omega_i$ of each driver is sampled from a normal distribution $\mathcal{N}(1, 0.2^2)$, but the optimizer always forecasts future drivers with $\omega=1$. Given $\omega_i$, we can calculate the number of vacancies in the RC queue by \eqref{eq:vrobo} and \eqref{eq:qrobo}. In the optimization model, the choice is deterministic: suppose the number of vacancies is $v$, the driver stays if $v > 0$ and leaves otherwise. While in the simulator, a sigmoid-like probabilistic model is used: the probability of staying is $(1 + a \exp{-bv})^{-1}$, where we use $a=2$ and $b=2$, so $P(\text{stay}) = 0.79, 0.33, 0.06$ when $v=1, 0, -1$ respectively.

\paragraph{Earlier/later departures} We consider the disturbance that PEVs may depart earlier or later than their registered departure time. Early departure may create an unsatisfied charging experience, and late departure may cause overstay if being assigned to FCs. To simulate this uncertainty, we make the actual departure time a random variable $\mathcal{N}(\widehat{t}_d, \sigma^2)$ where $\widehat{t}_d$ is the registered departure time of a session, and $\sigma$ is set as 15 minutes (and the actual duration is clipped to be 15 minutes if it is even shorter).

\bibliographystyle{IEEEtran}
\bibliography{Robo_main.bib}

\end{document}